\documentclass[prd]{revtex4}
\usepackage{amsmath}
\usepackage{amssymb}
\usepackage{graphicx}

\newcommand{\omLR}{\omega_{\rm LR}}

\begin{document}

\title{Black hole binary inspiral: Analysis of the plunge}

\author{Richard H.~Price} \affiliation{Department of Physics \&
  Astronomy, and Center for Advanced Radio Astronomy, University of
  Texas at Brownsville, Brownsville TX 78520}

\author{Sourabh Nampalliwar} \affiliation{Department of Physics \&
  Astronomy University of Texas at Brownsville, Brownsville TX 78520}

\author{Gaurav Khanna} \affiliation{Department of Physics, University
  of Massachusetts, Dartmouth, MA 02747}

\begin{abstract}
Binary black hole coalescence has its peak of gravitational wave
generation during the ``plunge,'' the transition from quasicircular
early motion to late quasinormal ringing. Although advances in
numerical relativity have provided plunge waveforms, there is still no
intuitive or phenomenological understanding of plunge comparable to
that of the early and late stages. Here we make progress in developing
such understanding by relying on insights of the linear mathematics of
the particle perturbation model for the extreme mass limit. Our
analysis, based on the Fourier domain Green function, and a simple
initial model, point to the crucial role played by the kinematics near
the ``light ring'' (the circular photon orbit) in determining the
plunge radiation and the excitation of QNR. That insight is then shown
to successfully explain 
results obtained for particle motion in
  a Schwarzschild background.
\end{abstract}

\maketitle

\section{Overview}\label{sec:overview}
\subsection{Background}

Interest in the coalescence of binary compact objects was motivated by
the development of gravitational wave (GW) detectors, and the need for
a better understanding of the endpoint of binary merger\cite{NRreview}. 
Especially in the case that such mergers end in black hole formation these 
might be the most powerful, and certainly the most interesting sources of
detectible GWs. Beyond that main motivation, there are several other 
reasons such mergers are of interest to the physics research community; 
see Ref.\cite{lrr2015} for a detailed review. From the outset it was 
understood\cite{earlyref} that binary coalescence, driven by GW radiation, 
could be viewed as having three stages.  The first is the gradual inspiral 
of the objects in orbits, most probably nearly circular orbits, gradually 
decreasing in orbital energy, and hence in radius, driven by the radiation 
reaction to the GWs being emitted. This first stage can be understood as 
a quasi-Newtonian process with GW radiation reaction treated as a time 
averaged loss of orbital energy. Considerable accuracy could be achieved 
by using post-Newtonian (slow motion, weak field) approximations\cite{pNref}.

The last stage of binary coalescence is the oscillation of the black
hole formed by the coalescence. These oscillations, after initial
transients, are the ``quasinormal ringing'' (QNR) of the final
hole. It has been known since the early 1970s\cite{QNRrefs} that these
QNRs constitute a spectrum of complex (i.e., damped) frequencies
characteristic of the mass and spin of the final black hole.

The second stage, the middle stage in the binary coalescence, may
usefully be called the ``plunge.'' It is the end of the early gradual
inspiral, and determines the magnitude and phase of the QNR. Most
important, this relatively short transition is the stage during which
the major part of the GW energy is
radiated\cite{PlungeImport}. Ironically, this most-important stage is
the part of the coalescence that is least understood.

For a wide variety of merger scenarios, the excitation of QNR {\em is}
known. The remarkable progress in numerical relativity in the past
decade\cite{numrelbreakthrough} has given detailed results for the GW
signals produced.  These results, however, do not give a qualitative,
phenomenological understanding of the plunge process, an understanding
at the same level as that we have for the early inspiral and late
QNR. In addition to the satisfaction that such an understanding could
provide, there are possible practical advantages. With an intuition
about the plunge, computational searches could be guided to
particularly interesting scenarios.  In this paper we attempt to give
a first step, an intersting first step, towards the understanding
of the plunge.

\subsection{Introduction and Summary}

We attempt here to give a simple understanding, more precisely a
simple view of the generation of plunge radiation and the excitation
of QNR. The ultimate goal is a set of guidelines that could allow a
complete quasi-analytic description of merger. It should be noted in
this connection that the difficulty posed by the plunge, the end 
of inspiral and beginning of QNR, has been
recognized as a challenge in the effective one body
formalism\cite{DamourGopakumar06}.

The result we present is a very simple general picture: the plunge is
determined by the conditions of the binary associated with the ``light
ring,'' (LR) the location of photon orbits. This picture is supported both
by a mathematical reason that the picture should apply, and by a
variety of examples. The simple general picture, however, must be
viewed as somewhat speculative since it depends on several
approximations. By far, the most important of these is the particle
perturbation approximation. We have also limited our analysis to
nonrotating (Schwarzschild) black holes and, in some aspects of the
mathematical analysis, to simplified models. Despite these
approximations, we feel that the picture that emerges is very
compelling as a first step that can motivate and guide further
work. 

There has been considerable past
work\cite{Goebel,PoschlTeller,BuCoPr:2006,Berti:2007,Zim:2012,Dolan:2011}
reported on the connection between the LR and QN frequencies. Of
particular relevance to our current work is the analysis of the
multipole content of merger ringdown radiation and the relationship of
the start of ringdown to the passage of the particle through the light
ring\cite{particleLR}.  Our interest here is somewhat different, not
only to observe the connection in the radiative signal but to
understand its origin, and to understand the transition to the
ringdown and the peak of the radiation.

The approach presented in the sections below consists of the following
steps.  First, in Sec.~\ref{sec:FDGF} we introduce the particle
perturbation approximation and the idea of using the Fourier domain
Green function (FDGF) to understand, with simple mathematics, the
analytic connection between the motion of our point particle and the
excitation of plunge radiation and QNR. 
With this technique, the radiation is reduced to a single, relatively simple,
integral over the path of the particle.
The use of the FDGF, in fact, is one of the
elements of our hope that the plunge can be understood. The basis of
that hope is that the QN frequency is a pole in the FDGF.  This is true in very broad
generality, so the way in which a particle source ``excites'' a pole
should have significant universality. Excitation in a simple model may
therefore tell us a great deal about a much more complicated realistic
model. With this motivation we introduce, in this section, a model with
a very simple FDGF.  Even this simple model, however, is complicated
enough in its details to be distracting, so most of these details are
deferred to an appendix.

Next, in Sec.~\ref{sec:Cubic} we define a class of trajectories that
allow for simple analysis combined with sufficient flexibility to test
models. In this section, also, we show that the perturbations computed
from the FDGF, for the simple model, agree to high accuracy with
numerical evolution of initial data. We then, in
Sec.~\ref{sec:radial}, use the FDGF to investigate what it is about
the trajectory that determines the excitation of QNR.  Our
investigations in this section are limited to radial motions, since
our model is based on a scalar charged particle source of a scalar
field. For the scalar case, orbital motion modifies the source with a
multiplicative, time dependent angular factor.  By contrast, for the
gravitational and electromagnetic case, the source term has extra
additive terms for angular motion. Our scalar model, therefore, will
not give results that are representative of gravitational wave
generation, it does however give interesting insights into the
excitation of QNR and to the plunge radiation.

Our conclusions from Sec.~\ref{sec:radial} are extended and tested in
Sec.~\ref{sec:Schw}, where we show them to be valid for trajectories
in the Schwarzschild background. In this section we include
investigations both of radial motion and motion with angular orbital
motion and we show in this section that the plunge radiation and QNR
are determined by the angular as well as the radial motion of infall.

In Sec.~\ref{sec:discussion}, we give a viewpoint on why the LR plays the
important role in the generation of QNR. We conclude in
Sec.~\ref{sec:conc} and relate the work presented to directions for
further investigations.  Some details of the complete FDGF are given
in Appendix~\ref{sec:AppA}.

We use units in which $c=G=1$ and other conventions of the text by
Misner, Thorne and Wheeler\cite{MTW}.  The following acronyms will be
used throughout the paper: QN=quasinormal; QNR=quasinormal ringing;
LR=light ring; FDGF=Fourier domain Green function. We will typically 
use the capital $T$ to indicate particle time, and the lower case $t$
for other uses, in particular as part of the retarded time of received 
radiation.

\section{Particle Perturbation Approximation and Fourier Domain Green Function}\label{sec:FDGF}

\subsection{Particle perturbation model}
We start by considering a model of a perturbation field $\Phi$
generated by particle motion in a Schwarzschild background.  In
effect, this particle perturbation model is justified by the fact that
it applies in the limit of an extreme ratio of the masses of the
binary compact objects\cite{LoNaZlCa:2010,SpCaOtScWi:2011,Berti:2010,NaZlLoCa:2011}. 
But an important question is whether phenomena in the extreme mass limit 
are qualitatively similar to the phenomena for comparable mass binaries.  
Many examples show that this does seem to be the case. One such example 
is the analysis of the antikick phenomenon found with numerical relativity 
computations for comparable mass\cite{paper1,paper2,SchnittmanEtAl}. With 
particle perturbation analysis \cite{SundararajanEtAlI,paper1,paper2} it 
has been shown that this paradoxical phenomenon has a simple underlying 
explanation. The current paper uses the history of such successes as the 
reason to look for particle perturbation modelling as an appropriate step
toward understanding the plunge and the excitation of QNR.

The great advantage of particle perturbation theory is, of course,
that the equations describing the fields are linear, and allow the
insights of a Fourier transform and, for a spherically symmetric background,
of multipole decomposition.

For greatest simplicity we take our model to be based on a scalar
perturbation field, $\Phi$, coupled to a scalar-charged particle.  We
start the analysis by decomposing $\Phi$ into spherical harmonics
$r^{-1}\sum_{\ell m}\Psi_{\ell m}Y_{\ell m}$ and by writing the
equation for the $\ell m$ multipoles as
\begin{equation}\label{eq:Schweq}
  \frac{\partial^2\Psi_{\ell m}}{\partial r^{*2}}
-\frac{\partial^2\Psi_{\ell m}}{\partial t^2}
-V_{\ell}(r^*)\Psi_{\ell m}=S_{\ell m}(r^*,t)\,.
\end{equation}
Here $r^*$ is the Regge-Wheeler\cite{RW57} ``tortoise coordinate''
defined in terms of the standard Schwarzschild areal coordinate $r$, by
\begin{equation}\label{eq:rstardef}
  dr/dr^*=1-2M/r\,.
\end{equation}
The crucial feature of the $r^*$ coordinate is that it remaps the
semi-infinite scope ($2M$ to $\infty$) of the Schwarzschild areal
coordinate to $-\infty$ to $\infty$.

The ``curvature potential''\cite{curvpot}, $V_\ell(r^*)$ depends on
the type of field (scalar, electromagnetic, gravitational
perturbation)\cite{andmath}.  What is common to all these potentials
is that for $r^*\gg M$ the potential has the form of a centrifugal
potential,
\begin{equation}
  V_\ell(r^*)\longrightarrow\frac{\ell(\ell+1)}{r^{*2}}\quad\mbox{for $r^*\gg M$}\,,
\end{equation}
and that as $r^*\longrightarrow-\infty$ the potentials fall off
exponentially in $r^*$, that is, as $(1-2M/r)$. The transition between
the flat spacetime centrifugal potential and the dramatically
decreasing exponential form occurs around the ``light ring,'' $r=3M$,
the location of circular photon orbits.

For the problem in which we are interested, the source term $S_{\ell
  m}(r^*,t)$ is a point particle moving in the equatorial
($\theta=\pi/2$) plane with radial position $r^*=F(t)$. The nature of
the source term, like the specifics of the potential, depends on
details of the model. The source term may contain a delta function, or
derivatives of delta functions, or sums of such terms. Our
paradigmatic simplest case will be the scalar problem, with a source
term that contains only a delta function, not its derivatives. In this
case the source will have the form
\begin{equation}
  S_{\ell m}(r^*,t)=f(t)\delta\left(r^*-F(t)\right)\,.
\end{equation}
Here $f(t)$, which in general may depend on $\ell,m$, can be used to
represent, e.g., time dependence of scalar charge or, as shall be
discussed later, orbital motion. This factor will almost always be taken
as $f=1$, but will be useful in Sec.~\ref{sec:Cubic}
for comparing numerical results.

We now change notation in two ways. First, we drop the $\ell,m$
indices which will be clear from the context of the models to
follow. Second, we will want to generalize the background rather than
be limited to that of the Schwarzschild geometry.  
Our generalization, in fact, focuses on Eq.~(\ref{eq:Schweq})
as a wave equation in one spatial dimension.
To emphasize this, we
replace $r^*$ by the coordinate $x$, with the understanding that it
ranges from $-\infty$ to $+\infty$. Equation (\ref{eq:Schweq}), then
is replaced by
\begin{equation}\label{eq:xeq}
\frac{\partial^2\Psi}{\partial x^{2}}
-\frac{\partial^2\Psi}{\partial t^2}
-V(x)\Psi=f(t)\delta\left(x-F(t)\right)\,.
\end{equation}

\subsection{The Fourier Domain Green Function}

We define the time domain Green function $G(x,a;t-T)$ by the
differential equation
\begin{equation}\label{eq:defG}
  \frac{\partial^2G}{\partial x^2}- \frac{\partial^2G}{\partial t^2}
  -V(x) G=\delta(x-a)\delta(t-T)\,,
\end{equation}
and the same boundary conditions that apply to Eq.~(\ref{eq:Schweq}):
outgoing radiation (i.e., propagating to larger $x$) at
$x\longrightarrow\infty$, and ingoing radiation (i.e., propagating to
lower $x$) at $x\longrightarrow-\infty$.  Since
\begin{equation}
  \delta(x-F[t])=\int_{-\infty}^\infty\int_{-\infty}^\infty
  \delta(a-F[T])\delta(x-a)\delta(t-T)\,da\,dT\,,
\end{equation}
it follows that the solution to Eq.~(\ref{eq:xeq}) is
\begin{equation}\label{eq:PsifromG}
  \Psi(t,x)=\int_{-\infty}^\infty\int_{-\infty}^\infty
G(x,a;t-T) f(T)\delta(a-F[T])\,da\,dT\,.
\end{equation}

Next we introduce the Fourier domain Green function (FDGF) ${\cal G}$
by
\begin{equation}\label{eq:GandcalG}
  G(x,a;t-T)=\frac{1}{2\pi}\int_{-\infty}^\infty e^{-i\omega(t-T)} 
{\cal G}(x,a;\omega) d\omega\,,
\end{equation}
where (from Eq.~(\ref{eq:defG})), ${\cal G}$ satisfies
\begin{equation}\label{eq:FDGF}
 \frac{\partial^2{\cal G}}{\partial x^2}+\left(\omega^2-V(x)\right) {\cal G}=\delta(x-a).
\end{equation}
We can put together Eqs.~(\ref{eq:PsifromG}) and (\ref{eq:GandcalG})
to get 
\begin{equation}\label{eq:PsifromcalG}
  \Psi(t,x)=\frac{1}{2\pi}\int_{-\infty}^\infty\int_{-\infty}^\infty\int_{-\infty}^\infty
  e^{-i\omega(t-T)} {\cal G}(x,a;\omega)
  f(T)\delta(a-F[T])\,d\omega\,da\,dT\,.
\end{equation}

\subsection{The FDGF for a simple model}\label{subsec:FDGFsimple}
The FDGF for the Schwarzschild curvature potential has a complicated
analytic structure. The infinite set of quasinormal frequencies must
appear, and -- even worse -- there is a branch point at $\omega=0$
associated with the late-time power-law tails\cite{tails}.  This
complexity is a disadvantage for intuitive insights, and adds
significant difficulty to evaluations of Green function solutions. For
that reason we illustrate some of the features of our analysis with a
model that has a very simple Green function. The truncated dipole
potential (TDP)\cite{tdp} is defined by
\begin{equation}\label{eq:TDPpot}
  V=\left\{
    \begin{array}{cl}
    \ell(\ell+1)/x^2=  2/x^2&\ {\rm for}\ x>x_0\\0&{\rm for}\ x<x_0 
    \end{array}
\right.\,.
\end{equation}
In this simple model, the sharp edge at $x=x_0$ plays the role of the
steep dropoff of the Schwarzschild potential for $r$ less than around
$3M$.  With the substitution of Eq.~(\ref{eq:TDPpot}) for the
Schwarzschild curvature potential, the solutions to
Eq.~(\ref{eq:FDGF}) take an elementary form both for $x<x_0$ and
$x>x_0$. Matching conditions then lead to a simple polynomial equation
to be solved for the QN frequencies.

There are disadvantages to making the substitution of the TDP for the
Schwarzschild potential, both obvious and not. The obvious
disadvantage is that the substitution raises the question whether the
insights provided will be applicable to the Schwarzschild
background. A partial answer to this question is the observation that
the phenomenon of complex frequencies occurs in the TDP (and in other
truncated multipole potentials). It is reasonable to infer that 
the excitation of the TDP and the Schwarzschild potential 
will show {\em qualitatively} similar dependences on particle motions.
The real answer to this question, however, is found in the models of
Sec.~\ref{sec:Schw}; the insights from the TDP model  {\em do}
seem to give correct insights into the nature of plunge radiation and
QNR in Schwarzschild models.

The not-so-obvious disadvantages of the TDP model are, however, worth a
consideration.  One of these disadvantages is the ``quality factor''
of the QN modes, the ratio of the real (oscillatory) part of a QN
frequency to the imaginary (damping) part.  For the least damped of
the Schwarzschild quadrupole modes, the modes of prime importance to
GW astrophysics, this ratio is 4.2, while for the single TDP pair of
modes it is 1.0.  Due to the low quality factor, QNR in TDP radiation
does not have the same lightly damped sine wave appearance as for QNR
in Schwarzschild models. Worse than this visual inconvenience is the
fact that the damping part of the QNR for TDP models is smaller than
that for the Schwarzschild models. (More specifically, the TDP damping
frequency is $1/(2x_0)$ or $1/6M$ if we associate $x_0$ with $3M$. The
Schwarzschild damping frequency $0.088965/M$ is about half of that.)
We will argue that QNR is determined by the features of the orbit
within a damping time of crossing the LR. The longer damping
time for TDP than for Schwarzschild, means that this principle of
LR importance is tested more stringently in the TDP models
than in Schwarzschild models.

It would be ideal, of course, to have a model for the potential term
that has all the features we would like, but for which the FDGF has an
elementary form. A sharp cutoff at some $x_0$, as in
Eq.~(\ref{eq:TDPpot}), guarantees that for $x<x_0$ the Green function
has a simple elementary form. The challenge is to find the right form
of $V(x)$ for $x>x_0$. One such possibility is the P\"{o}schl-Teller
potential\cite{PoschlTeller}, sech$^2x$, but this potential does not
approach the flat spacetime centrifugal potential at large $x$ and
hence is a questionable subsitution.  An obvious question is why we do not
use a 
higher order truncated multipole potential, i.e.,
$\ell(\ell+1)/x^2$, with an integer $\ell$ larger than unity. For all
such potentials the Green function has only elementary functions.  As
$\ell$ increases, however, the Green function -- though elementary--
becomes more and more complicated. Furthermore, the quality factor of
the least damped QN mode increases slowly with increasing $\ell$. We
would have to go to $\ell$ of order 10 to get a ratio of damping to
oscillation that is qualitatively similar to QNR in the Schwarzchild
geometry\cite{TMPelsewhere}.

The complete FDGF for TDP, and
its derivation, are given in Appendix~\ref{sec:AppA}. The explicit
expressions for the FDGF depend on the relative location of $x$,
$x_0$, and particle position $a$, but all expressions are simple. As an
example, for $a<x_0<x$ the expression is
\begin{equation}\label{eq:TDPFDGF}
 {\cal G} =-\,
 \frac{i}{2}\frac{e^{-i\omega(a-x)}\omega}{(\omega-\omega_1)(\omega-\omega_2)}
 \left(1+\frac{i}{\omega x}\right)\,,
\end{equation}
where $\omega_{1,2}$ are the TDP QN frequencies
\begin{equation}\label{eq:TDPomegas}
\omega_1,\omega_2=-\frac{i}{2x_0}\pm\frac{1}{2x_0}  \,,
\end{equation}
which show up, as expected, as poles in the FDGF.

For a known trajectory $x=F(T)$, we can find the solution to
Eq.~(\ref{eq:xeq}), for $x>x_0$, at all times, by using ${\cal G}$ of
Eq.~(\ref{eq:TDPFDGF}) for $a<x_0$, and the Green function (in
Eq.~(\ref{eq:convenient})) for $a>x_0$.  To do this we introduce
notation for times of importance $T_{\rm cross}$ and $T_6$, important
times along the trajectory. (The notation $T_6$ arises from the
step-by-step development of the TDP FDGF in Appendix~\ref{sec:AppA}.)
These times, and the integration variables $\theta$ and $\xi$, are
defined by
\begin{equation}\label{T6Tcross}
   F(T_{\rm cross})=x_0 \quad\quad    T_6-F(T_6)=t-x
\end{equation}
\begin{equation}\label{eq:defsthetaxi}
  \theta=(t-T+F(T)-x)/(2x_0)\quad\quad\xi={(t-T-x-F(T)+2x_0)}/{(2x_0)} \,.
\end{equation}
We take $f(t)=1$; the solution is then
\begin{displaymath}
  \Psi=\frac{x_0}{2}\,\int_{T_{\rm cross}}^{T_6}e^{-\theta}
  \left[\frac{1}{x_0}\left(\sin\theta-\cos\theta\right)-\frac{2}{x}\sin\theta\right]\,dT
\end{displaymath}
\begin{equation}\label{eq:Psiintegrals}
    -\,\frac{1}{2}\,\int_{-\infty}^{T_{\rm cross}}
\,e^{-\xi}\left[
-(\cos{\xi}+\sin{\xi})+2x_0\left(\frac{1}{F(T)}+\frac{1}{x}\right)\cos{\xi}
-\frac{2x_0^2}{xF(T)}\left(\cos{\xi}-\sin{\xi}\right)
\right]\,dT\,.
\end{equation}
The second integral extends to $T_{\rm cross}$, the time at which the
particle, moving inward, passes the potential ``edge'' at
$x=x_0$. This integral represents the contributions to the field from
the pre-crossing motion; the first integral gives the contribution for
the motion subsequent to the edge passage.

\section{Test Trajectory}\label{sec:Cubic}

To investigate the generation of radiation, we must choose particle
trajectories for the Schwarzschild background, for our TDP, or for the
other models that we will introduce below. There are several features
we would like in a family of trajectories. First, we would like all
trajectories in the family to conform to the analog of the physical
constraints on trajectories in the Schwarzschild background. The most
fundamental of these constraints is that $dr^*/dt\rightarrow-1$ as the
particle approaches the horizon. For our trajectory family to be used
with Eq.~(\ref{eq:xeq}) we therefore require $dx/dt\rightarrow-1$ as
$x\rightarrow-\infty$.

We also require that the trajectories belong to a family with at least
two parameters. In this way, we can compare the radiation for a range
of trajectories with one feature of the trajectories fixed. It is in
this way that we will argue that the plunge radiation and QNR are not
sensitive to the nature of the trajectory away from the edge.

A third criterion for a family of trajectories is one of convenience.
As a check on details, it is useful to compare the computation of
$\Psi$ done with the FDGF integrals, like those in
Eq.~(\ref{eq:Psiintegrals}), to the solution found by evolving initial
data in the standard manner used in numerical relativity.  For such
evolution, we need initial data, and the most convenient initial data
is static initial data, i.e., the fields that would exist at time zero
if the particle had been sitting in a fixed position forever, and only
started to move at time zero. Note that the motion before time zero is
needed not only for evolution codes, but is also needed for the Green
function solution, since the integrals, e.g., the second integral in
Eq.~(\ref{eq:Psiintegrals}), require a specification of the particle
position back to the infinite past. For static initial data, both
$\partial\Psi/\partial t$ and $\partial^2\Psi/\partial t^2$ must
vanish at time zero. This in turn means that for our trajectories both
the velocity and acceleration must be zero at time zero.

A desideratum, though not a requirement, is that the family of
trajectories leads only to simple closed form analysis. The following
two-parameter family for the radial motion, our ``cubic trajectory''
family, satisfies this as well as the three criteria above:
\begin{equation}\label{eq:FofT}
  F(T)=\left\{
  \begin{array}{cl}
    a_0+\tau-\left(T^3+\tau^3\right)^{1/3}&T>0\\a_0&T<0\,.
  \end{array}
  \right.       
\end{equation}
The two parameters are $a_0$, the initial $x$ position of the
particle, and $\tau$, the timescale on which the particle
accelerates. For $T>0$, both the velocity $v$ and acceleration $a$
(not to be confused with ``$a$'' used in the previous section as a
proxy for $F(t)$) are
\begin{equation}
  v=-T^2\left(T^3+\tau^3\right)^{-2/3}\quad\quad\quad 
a=-2T\tau^3\left(T^3+\tau^3\right)^{-5/3}\,,
\end{equation}
and both go to zero as $T\rightarrow0$.

The cubic trajectory can be inverted in closed form to give $T$ in
terms of $F$.  What is actually required for computations are the
solution for $F(T)\pm T$. In the $F(T)- T$ case  we must solve
\begin{equation}
  T+\left(\tau^3+T^3\right)^{1/3}=\rho\,,
\end{equation}
where $\rho$ is a known function of $t,x,a_0,\tau$. 
The cubic equation that results has  the closed
form solution
\begin{equation}
  s\equiv\sqrt{\frac{\tau^6}{16}+ \frac{\rho^6}{64}\;}\quad\quad
  A\equiv\left(-\frac{\tau^3}{4}+s\right)^{1/3}\quad\quad
  B\equiv-\left(\frac{\tau^3}{4}+s\right)^{1/3}\quad\quad
  T=A+B+\frac{\rho}{2}\,.
\end{equation}
In the $F(T)+ T$ case,
the problem requires solution 
of 
\begin{equation}
  T-\left(\tau^3+T^3\right)^{1/3}=-\sigma\,,
\end{equation}
where $\sigma$ is a known function of the trajectory parameters. By transposing one of the terms from the left
to the right, and raising both sides to the third power, the $T^3$
terms cancel, leaving a quadratic equation which has the solution
\begin{equation} 
T=\frac{1}{2}\left(
-\sigma+\sqrt{\frac{4\tau^3}{3\sigma}-\frac{\sigma^3}{3}\;}
\right) \,.
\end{equation}

  \begin{figure}[h]
  \begin{center}
  \includegraphics[width=.4\textwidth ]{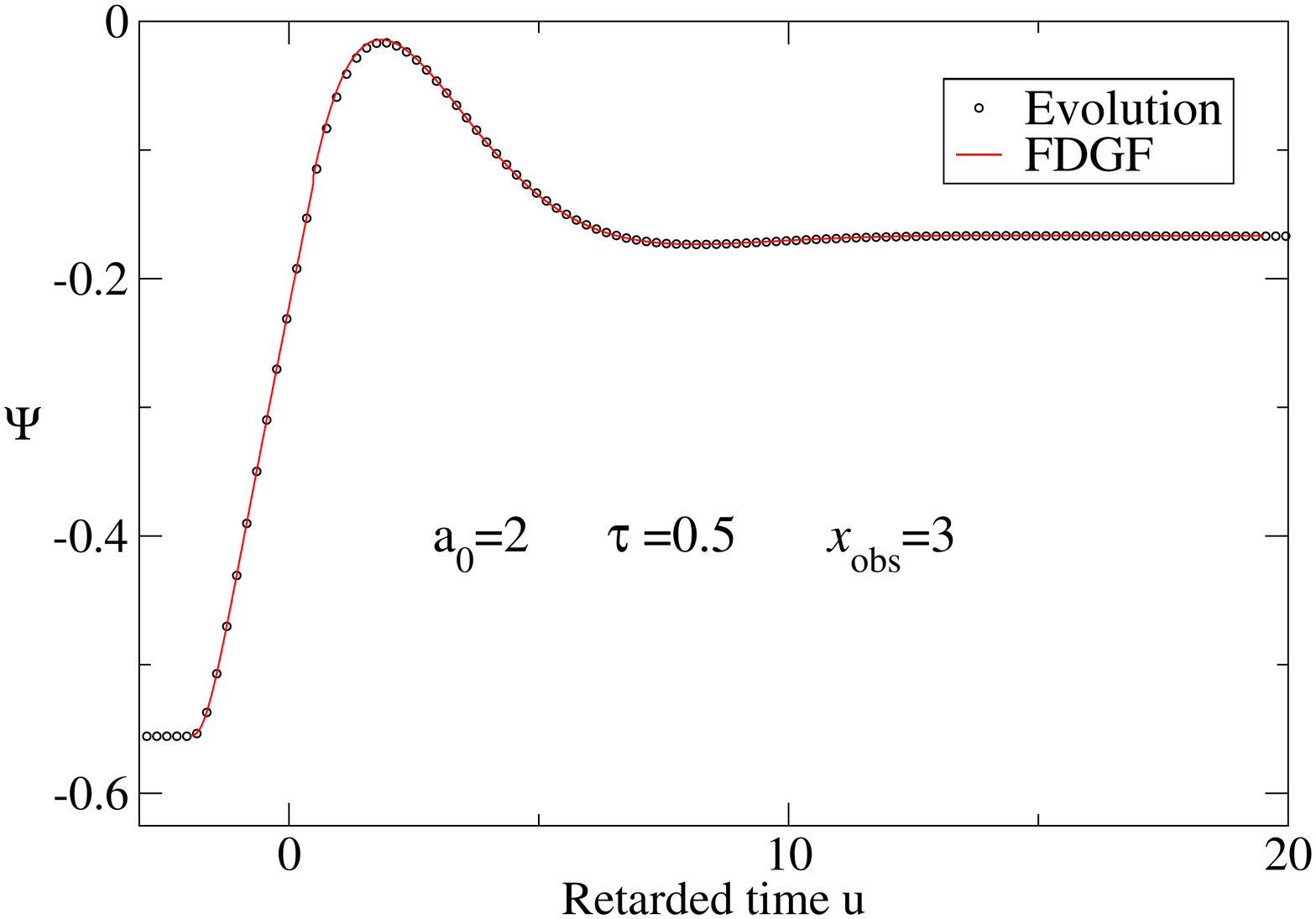}
  \includegraphics[width=.4\textwidth ]{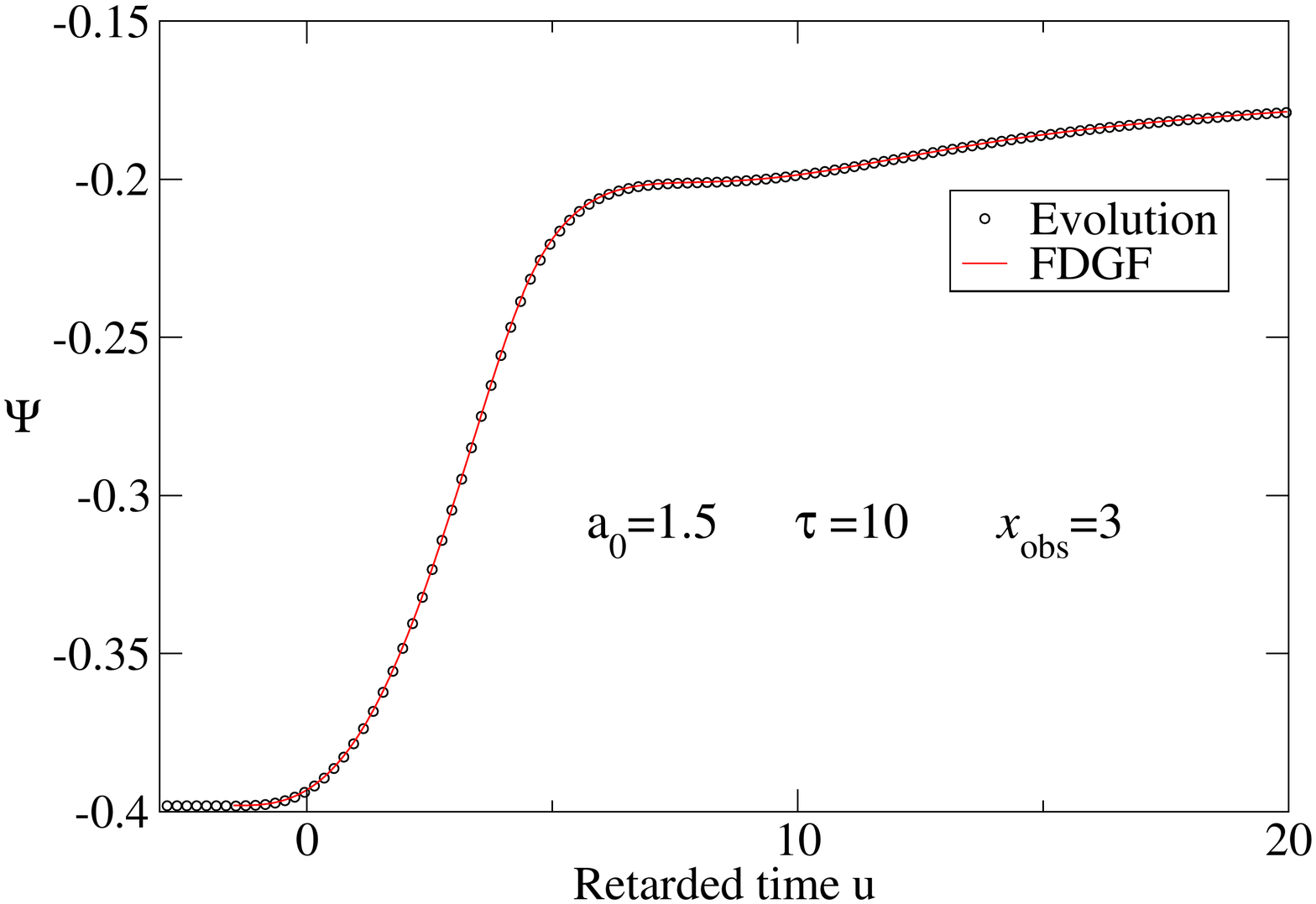}
  \caption{Comparison of computations with the Fourier domain Green
    function and with numerical evolution for radial infall. 
The scalar TDP test field $\Psi$ is 
 given as a function of retarded time $u$.
All values are given in 
units of the edge location $x_0$. The particle generating the field moves
    according to the cubic trajectory of Eq.~(\ref{eq:FofT}) with the
    parameters shown in each plot.  The particle has been stationary
    at $a_0$ prior to $T=0$.  }
  \label{fig:radialcomparo}
  \end{center}
  \end{figure}

Figure \ref{fig:radialcomparo} shows the comparison between numerical
evolution 
of the partial differential equation in Eq.~(\ref{eq:xeq}),
and the evaluation of the FDGF, for radial infall, with $f(t)=1$,
starting from stationary initial data at $t=0$ and following the cubic
trajectory of Eq.~(\ref{eq:FofT}).  
In this figure, and in those that follow for the TDP, the scaling is 
set by $x_0$, in a manner parallel to the scaling set in the Schwarzschild
geometry, in geometrized
($c=G=1$) units, by the mass parameter $M$. Thus, for
example, a retarded time $u=50$ is understood to mean $u=50x_0$ and an
acceleration $a=0.01$ is understood to mean $a=0.01/x_0$.
Note that $\Psi$ itself is also proportional to $x_0$.

The FDGF integrals were carried out using a straightforward Simpson's 
rule scheme.
The evolution scheme is the well-known second-order, Lax-Wendroff
finite-difference scheme for second-order 1+1 linear hyperbolic
equations; the delta function is represented by a narrow Gaussian, and
the boundary conditions are taken to be no incoming waves from
$x=\pm\infty.$
The TDP field $\Psi$, computed, or ``observed,'' (i.e., numerically
extracted) at $x=3$ is shown as a function of retarded time
$u=t-x$. (The choice of a relatively small value of the extraction
radius means that the results test all the terms in the TDGF, not only
the terms in the radiation zone.) Although the results shown here are
limited to the spacetime region $x>x_0$, and $x>F(T)$, computations in
other regions showed the same excellent agreement.

For a scalar charged particle falling with angular motion
$\phi=\tilde{\phi}(t)$, the source term on the right of
Eq.~(\ref{eq:defG}) has an additional factor of
$\delta(\phi-\tilde{\phi}(T))$.  (Though angular motion in our scalar
model is not representative of the change in the source for
gravitational waves, it remains a useful check on the agreement of the
FDGF and evolution.)  To compute $\Psi(t,x,\phi)$ for the $\ell=m=1$
multipole moment we must insert $f(t)=\exp{(\pm
  i(\phi-\tilde{\phi}(T))}$ into Eq.~(\ref{eq:PsifromcalG}).  To check
that the FDGF integrals and numerical evolution agree for orbital
motion, we take $\phi=0$, and we extract the real part of $\Psi$,
i.e.\,, we insert $\cos{\tilde{\phi}(T)}$ in the integral.  In order
to have the angular, as well as radial motion start with zero
acceleration, we take the angular motion to be
\begin{equation}\label{eq:phiofT}
  \tilde{\phi}(T)=\phi_0\left[
    \frac{\left(T^3+\tau^3\right)^{1/3}-\tau} {\left(T^3_{\rm
        cross}+\tau^3\right)^{1/3}} \right]\,.
\end{equation}
With this choice, orbital motion starts at $\phi=0$ with zero angular
velocity and angular acceleration.

  \begin{figure}[h]
  \begin{center}
  \includegraphics[width=.3\textwidth ]{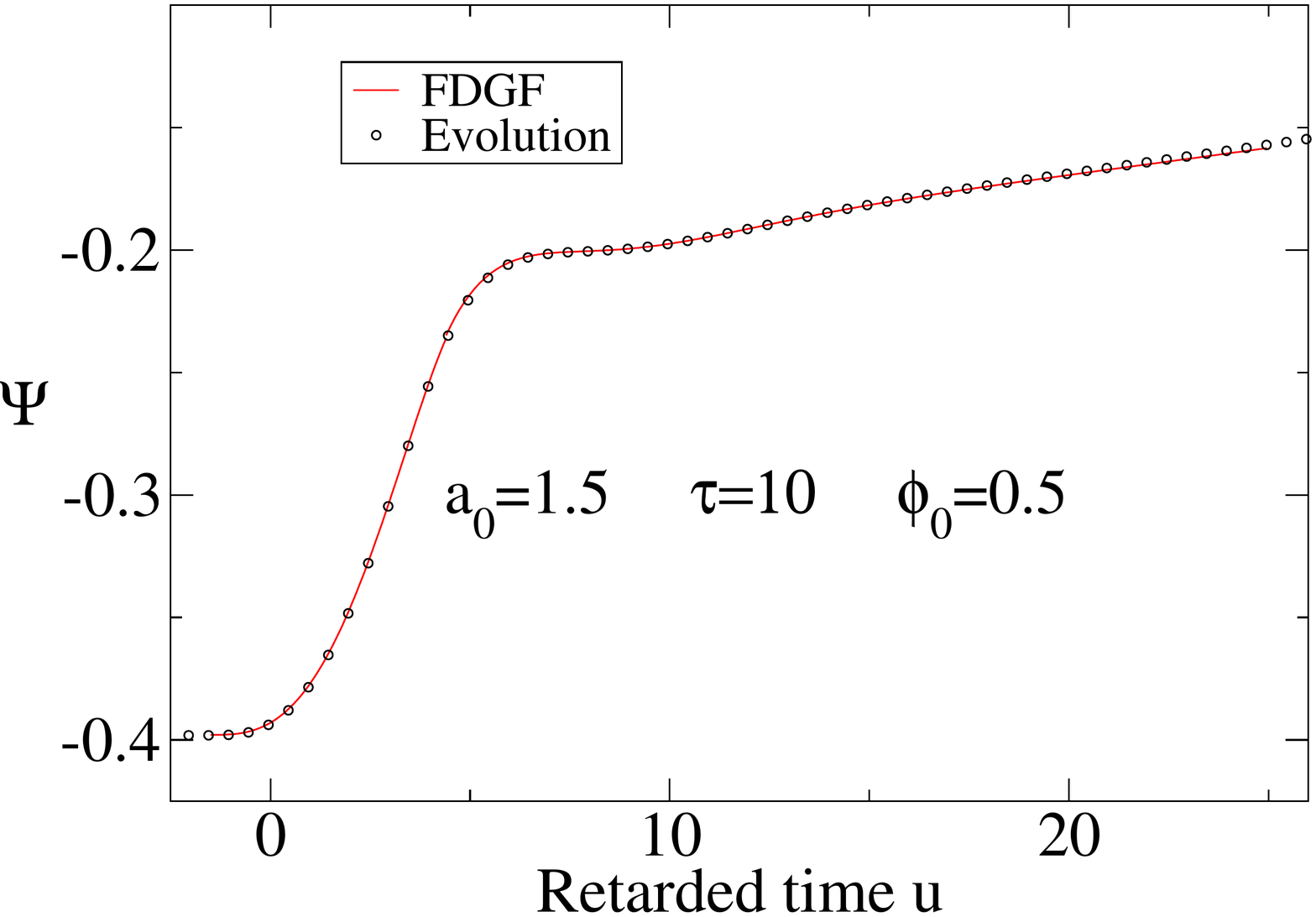}
  \includegraphics[width=.3\textwidth ]{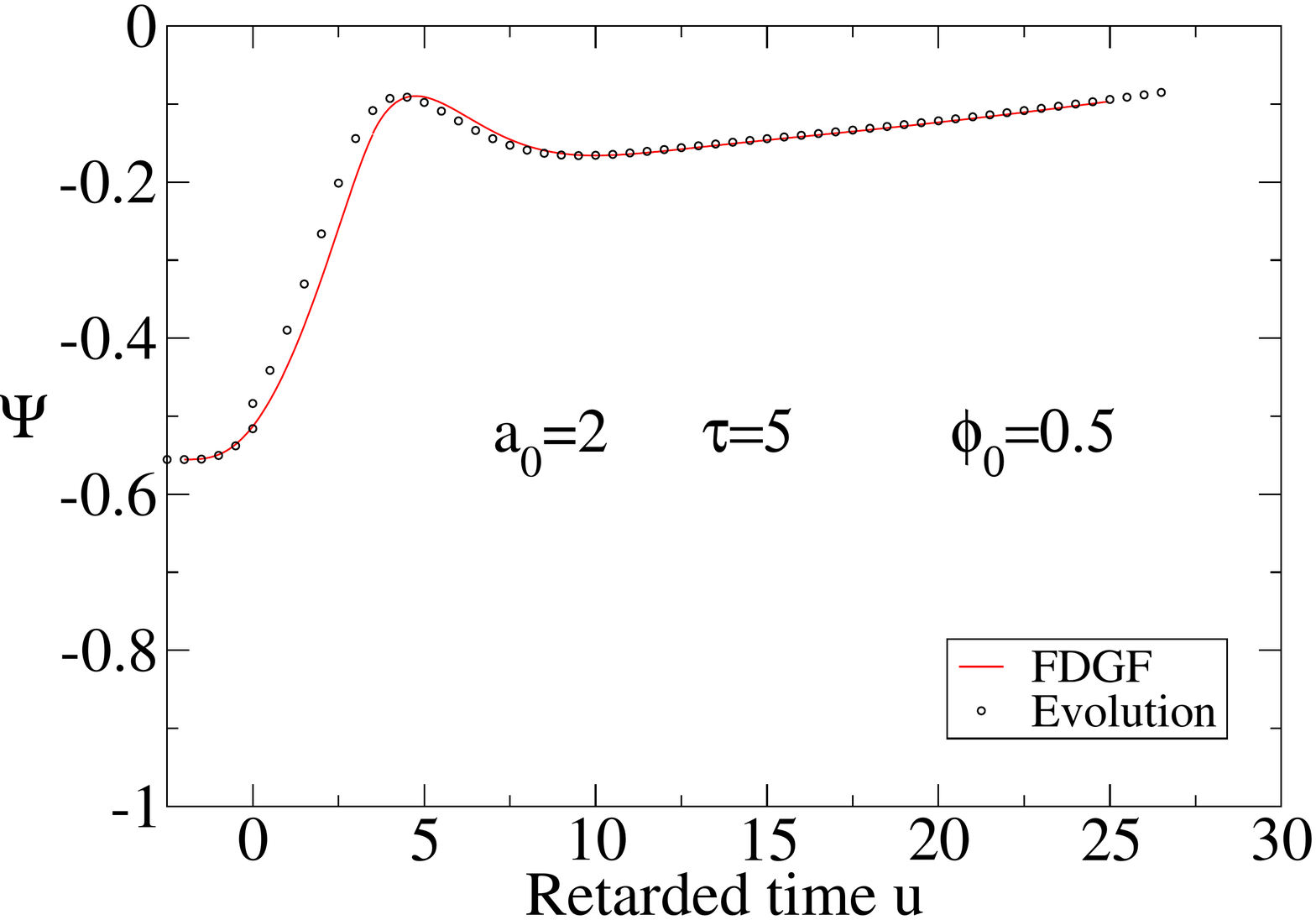}
  \includegraphics[width=.3\textwidth ]{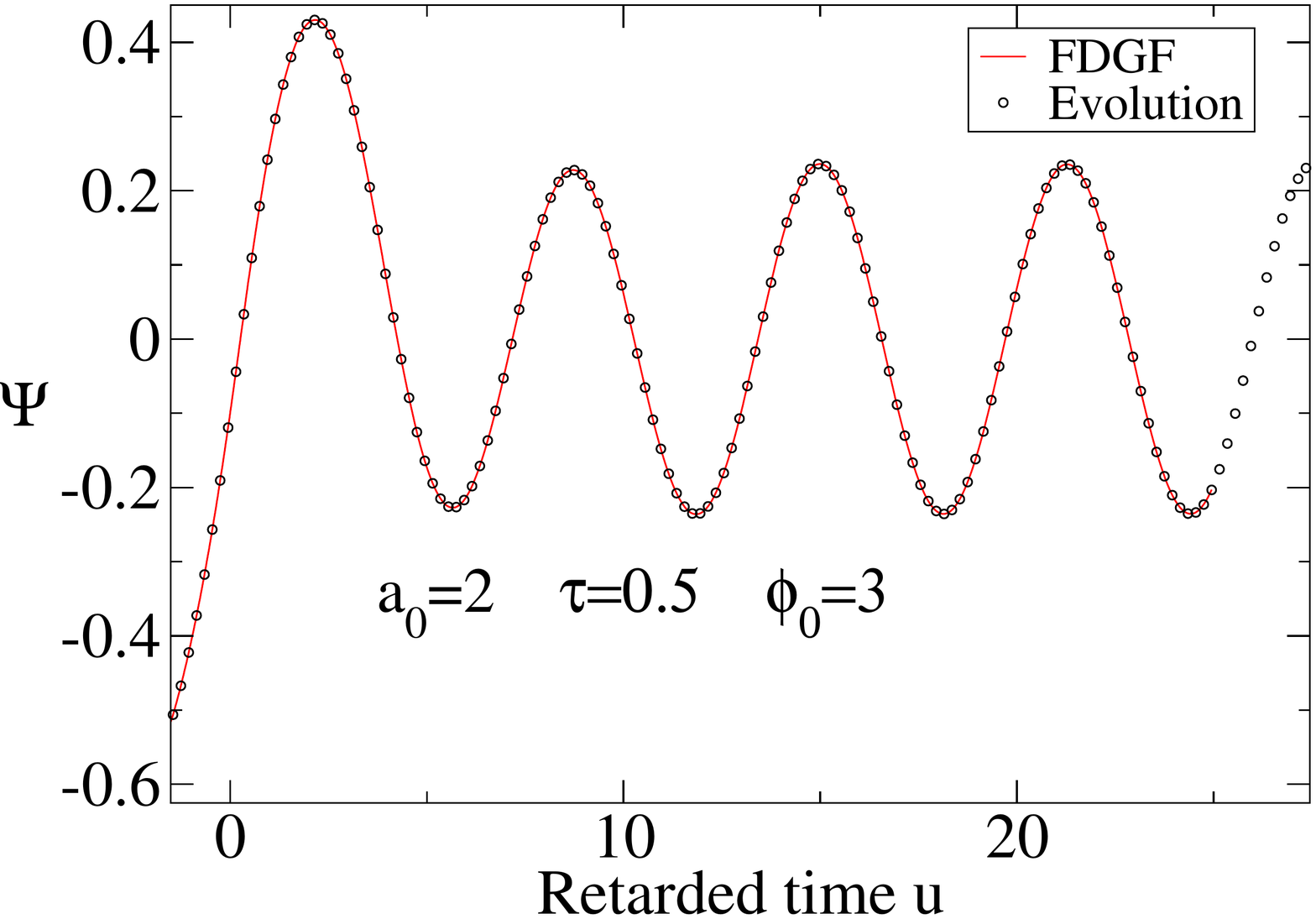}
  \caption{Comparison of computations with the Fourier domain Green
    function and with numerical evolution for orbiting infall. The
    radial motions are those described by Eq.~(\ref{eq:FofT}), with
    the trajectory parameters indicated in each plot; values are given in units of the edge
location $x_0$. The source has been
    modified with the angular factor $\cos{\tilde{\phi}(T)}$, as
    described in Eq.~(\ref{eq:phiofT}).  The examples give comparisons
    of three orbits that are qualitatively very different.}
  \label{fig:orbitalcomparo}
  \end{center}
  \end{figure}
Figure \ref{fig:orbitalcomparo} shows the comparison between numerical
evolution and evaluation of the FDGF for orbiting infall starting from
this stationary initial data at $t=0$. The $x$ position of the
particle follows the cubic trajectory of Eq.~(\ref{eq:FofT}) and the
azimuthal angle is given by Eq.~(\ref{eq:phiofT}).  Note that this
orbital motion is not physical. The angular motion is not compatible
with the conservation of angular momentum in the Schwarzschild
background which our TDP is meant to model. This, however, is not
relevant to the current purpose of testing the consistency of
evolution and FDGF computations.

\section{QNR from radial infall}\label{sec:radial}
\subsection{The mathematics of QNR excitation}
 To obtain insights into the plunge radiation and the  excitation of QNR in the TDP, we look at
 a typical scenario in which a particle starts outside the edge of the
 TDP (in terms of Eq.~(\ref{eq:FofT}), \(a_0 > x_0\)), crosses the
 edge (at a trajectory time \(T_{\rm cross}\)), and continues to fall
 in. The FDGF integrals in Eq.~(\ref{eq:Psiintegrals}) describe this
 scenario.  We can make this simple integral even simpler by
 considering only the radiation zone, i.e., by ignoring the part of
 $\Psi$ that falls as $1/x$.  This leaves us with
\begin{align}\label{eq:simpleint}
  \Psi= &\frac{1}{2}\,\int_{T_{\rm cross}}^{T_6}e^{-\theta}
\left(\sin\theta-\cos\theta\right)\,dT \nonumber\\
& -\,\frac{1}{2}\,\int_{-\infty}^{T_{\rm cross}}
\,e^{-\xi}\left[
\left (\frac{2x_0}{F(T)}-1\right )\cos{\xi}-\sin{\xi}
\right]\,dT\,.
\end{align}

We will call the integral involving \(\theta\)'s the ``late'' 
(after edge crossing) integral, and the one involving \(\xi\)'s the ``early''
(prior to edge crossing) integral and we start our analysis with the
late integral.  The expression for ${\theta}$ is illustrated in
Fig.~\ref{fig:T6} of Appendix~\ref{sec:AppA}, in which a complete
description is given of the FDGF. That figure, and the discussion
connected to it, show that the late integral expresses the influence
of the particle, consistent with causality, on the spacetime point at
retarded time $u$. What is of special importance is the ``damping,''
i.e., the falloff of influence dictated by the factor $e^{-\theta}$.

To understand the role of this damping factor, we note that for any
value of $u$, the damping factor is unity when ${\theta}=0$,
that is, the contribution is maximum from the point on the trajectory
labeled $T_6$ in Appendix~\ref{sec:AppA}. The time $T_6$, by
causality, is the latest time on the trajectory that can influence the
field at retarded time $u$.  Contributions from earlier points on the
trajectory, at time $T_6-\Delta T$, are damped by $e^{-\Delta\theta}$
with
\begin{equation}
  \Delta\theta=\left(\Delta T+\left[F(T_6-\Delta T)-F(T_6)\right]
\right)/2x_0\approx
\left(1+v_6\right)\Delta T/2x_0\,,
\end{equation}
where $v_6=|dF/dT|$, the speed of the trajectory at $T_6$, is
a positive quantity.

At the most basic level, this result shows that for the contribution
to the late integral at any value of $u$, the trajectory is relevant
only for a trajectory time less than or roughly comparable to the
damping time $2x_0$. It also shows that the only property of
the trajectory that matters is the velocity at $T_6$, unless the
velocity changes substantially over a time comparable to one damping
time, i.e., only if $|d^2F/dt^2|$ is comparable to
$2x_0|dF/dt|$.

A useful insight comes from changing the integration variable in the
first integral of Eq.~(\ref{eq:simpleint}) to ${\theta}$, so that the
integral becomes
\begin{equation}\label{eq:simpleintintheta}
  \Psi_\mathrm{late}=x_0\int^{{\theta}_{\rm cross}}_{0}e^{-\theta}
\frac{\left(\sin\theta-\cos\theta\right)}
{1+v({\theta})}\,d{\theta}\,,
\end{equation}
where $v({\theta})$ is $|dF/dT|$ expressed as a function of
${\theta}$ rather than $T$. If the velocity changes very little over
the range of $T$ during which most of the contribution to the integral
occurs, then we can approximate $v(\theta)=v_{\rm cross}$
in Eq.~(\ref{eq:simpleintintheta}) to get
\begin{equation}\label{eq:constvapprox}
\Psi_\mathrm{late}=\frac{x_0}{1+v_{\rm cross}}\,\int^{{\theta}_{\rm cross}}_{0}e^{-\theta}
\left(\sin\theta-\cos\theta\right)\,d\theta\,
=-x_0\frac{e^{-\theta_{\rm cross}}\sin\theta_{\rm cross}}
{1+v_{\rm cross}}\,.
\end{equation}
This means that the peak of $\Psi$ occurs at $\theta_{\rm cross}=\pi/4$, or at
\begin{equation}\label{eq:upeak}
  u_{\rm peak}-u_{\rm cross}=  u_{\rm peak}-T_{\rm cross}+x_0=\pi/4\,
\end{equation}
and that the value of $\Psi$ at the peak is
\begin{equation}\label{eq:psipeak}
  \Psi_{\rm late,peak}=-\frac{x_0\sqrt{2\;}e^{-\pi/4}}{\pi\left(1+v_{\rm cross}\right)}\,.
\end{equation}
We noted that only the velocity at crossing can matter. These results
show that, in fact, for the late integral the location of the peak of
the radiation is independent of that velocity, and the strength of the 
radiation is very insensitive to the velocity. A better statement, then,
is that the radiation described by the late integral is approximately independent of any feature of the trajectory.

\begin{figure}[h]
  \begin{center}
  \includegraphics[width=.26\textwidth ]{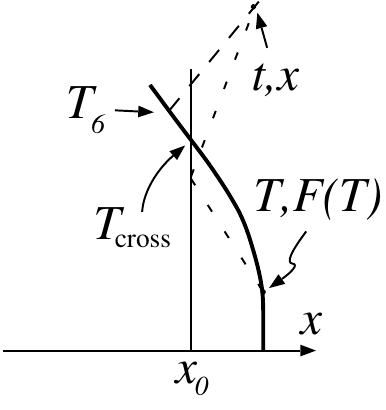}
  \caption{ A spacetime diagram showing the influence of the particle
    on the ``observation'' point at retarded time $u=t-x$. The dashed
    $45^\circ$ line indicates the direct, speed-of-light causality
    connecting the particle position at $T_6$ with the observation
    point, and showing that $T_6$ is the latest point on the particle
    world line that can influence observation at retarded time $u$.
    The  ``slower than light speed'' dashed lines indicate influence
    from earlier times via reflection from the edge at $x_0$.}
  \label{fig:T6plus}
  \end{center}
  \end{figure}

Though it is a reasonable approximation to take the velocity to be constant in the late
integral, it is not useful to take $F(T)$ to be a constant (in particular, to be $x_0$) 
in the second integral 
in Eq.~(\ref{eq:simpleint}).
Graphical insights, however, do follow from Figure~\ref{fig:T6plus}.
The figure  shows the two sources of the
radiation at the ``observation'' event at retarded time $u$. The
$45^\circ$, ``speed of light,'' line from $T_6$ to the observation
event illustrates that $T_6$ is the latest time on the world line that
can influence that event. This, then is the graphical explanation of
why the upper limit in the early integral in Eq.~(\ref{eq:simpleint})
is at $T_6$. The lower limit in this early integral is at $T_{\rm
  cross}$ because all of the integral in Eq.~(\ref{eq:simpleint})
arises from the TDGF for $F(T)<x_0$; for $T<T_{\rm cross}$ the
particle has not yet crossed that edge.

It is intuitively obvious  that the edge at $x_0$ must be involved in
QNR. 
This raises a natural question about the early integral in
Eq.~(\ref{eq:simpleint}). Since this integral is over only $T\leq
T_{\rm cross}$, its radiation features would seem not to be influenced by the existence
of the edge. How can it, therefore, contribute to QNR?
The question is answered graphically in
Fig.~\ref{fig:T6plus}; the early integral involves radiation that has
interacted with the edge via the dashed path in Fig.~\ref{fig:T6plus}.
We can see this in the form of ${\xi}$ in
Eq.~(\ref{eq:defsthetaxi}).  Notice that the total distance traversed along
the dashed line in Fig.~\ref{fig:T6plus} is the sum of the inward
distance $F(T)-x_0$ and the outward distance $x-x_0$. The total time
is $t-T$. It follows that ${\xi}$ is simply the total time minus
the distance traversed, so that causality requires that ${\xi}$
be greater than 0. But ${\xi}=0$ is equivalent to the $T=T_{\rm
  cross}$ upper limit of the the early integral.

It is important to understand that the ``slower than light''
propagation involved in the early integral is not a manifestation of
strong spacetime effects. Rather, it is associated with the fact that
we are working with a multipole decomposition. The influence of a
particle's multipole moments on the radiation signal at an event,
comes from all angles. The radial characteristics, then, are only an
indication of the limiting speed with which multipolar information
propagates.

\subsection{Models for cubic trajectories and TDP}
 As the first test of the QNR excitation principle explained above, we
look at a scalar charge following a radial trajectory in the TDP
background.  We use the radial trajectory function denoted by \(F(T)\)
and defined in Eq.~(\ref{eq:FofT}). From that equation and 
\begin{equation}
F(T_{\textrm{cross}}) = x_0 \nonumber
\end{equation}
we have that
\begin{equation}\label{eq:tlightringdefn}
  T_{\textrm{cross}} = \left [ (a_0 + \tau - x_0)^3 - \tau^3 \right ]^{1/3}\,.
\end{equation}
 Once $T_{\textrm{cross}}$ is known, the 
velocity and acceleration at crossing can be found from
\begin{align}
  v(T_{\textrm{cross}}) &= \left .\frac{d}{d\,T}F(T)\right
  |_{T=T_{\textrm{cross}}} =
  -\,\frac{T_{\textrm{cross}}{}^{2}}{(\tau^3+T_{\textrm{cross}}{}^3)^{2/3}},
  \\ a(T_{\textrm{cross}}) &= \left .\frac{d^2}{d\,T^2}F(T)\right
  |_{T=T_{\textrm{cross}}} =
  -\,\frac{2T_{\textrm{cross}}\tau^3}{(\tau^3+T_{\textrm{cross}}{}^3)^{5/3}}.
\end{align}
In all our models the velocity and acceleration are nonnegative.
For simplicity, we will use the absolute values of these quantities in
our discussions below, i.e., 
\begin{align}
  v_{\textrm{cross}} \equiv |v(T_{\textrm{cross}})|, \\
  a_{\textrm{cross}} \equiv |a(T_{\textrm{cross}})|.\label{eq:across}
\end{align}

\begin{figure}
  \includegraphics[width=0.48\textwidth]{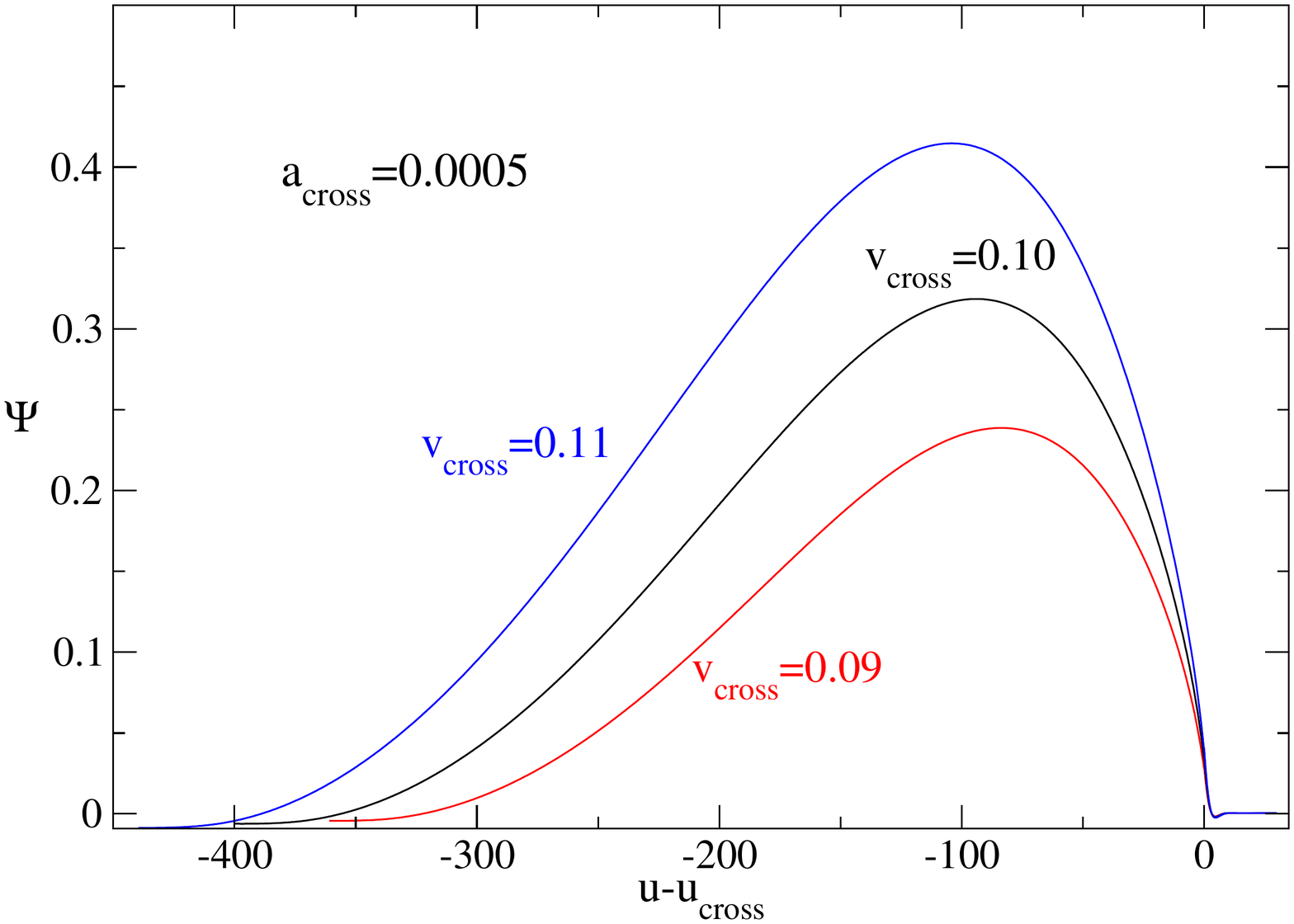}
  \includegraphics[width=0.48\textwidth]{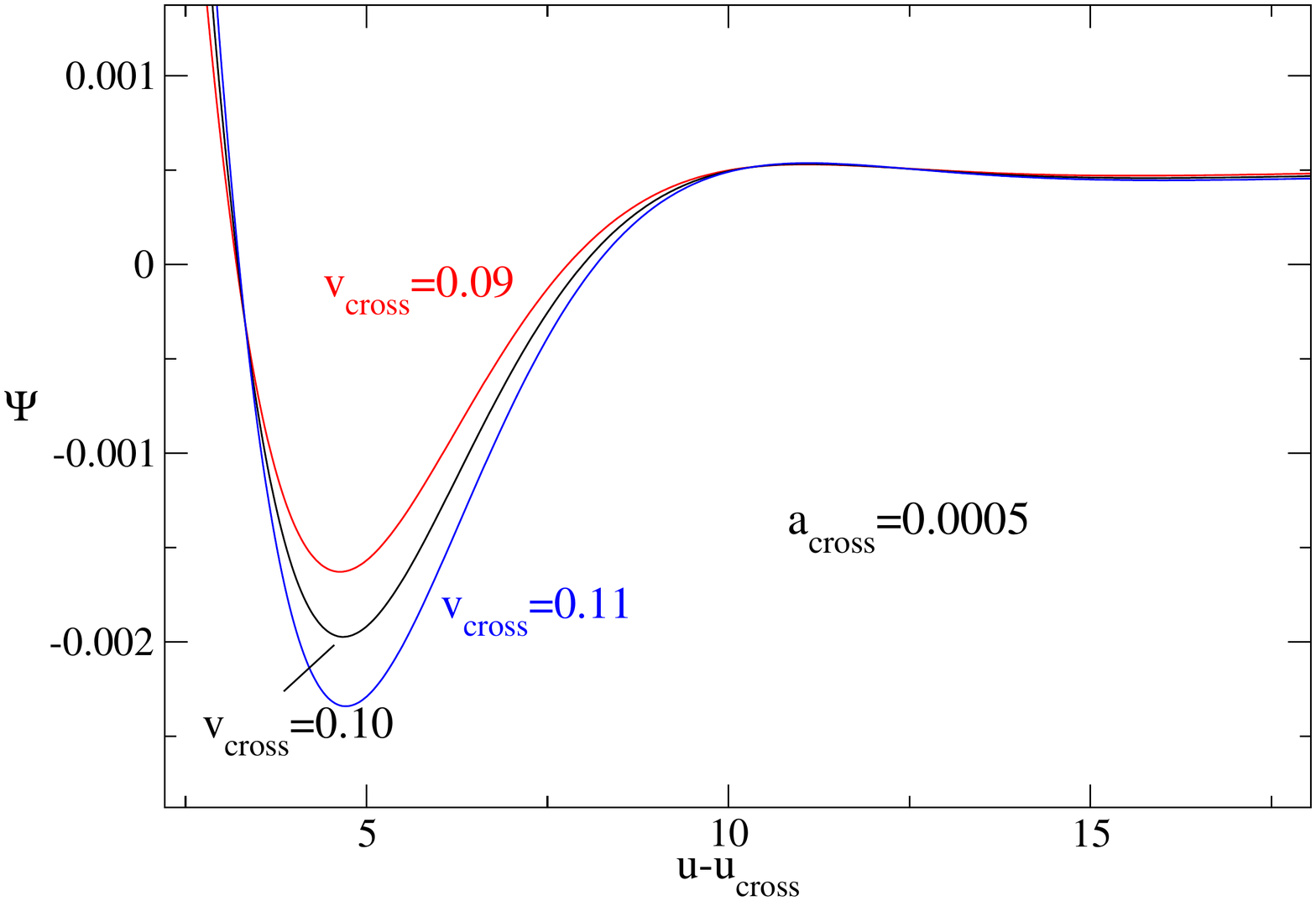}
\caption{The radiation due to a particle in radial infall in the TDP
  background. The horizontal axis is the retarded time after the edge
crossing value $u_{\rm
    cross}\equiv T_{\rm cross}-x_0$.
  The retarded time and $\Psi$
  are given in units of $x_0$; the acceleration is given in units of
  $1/x_0$. The panel on the right focuses on the region of the curves
with the peak QNR.}
\label{fig:comparo3vs}
\end{figure}

The radiation profiles for several radial infall trajectories, for
three different values of $v_{\rm cross}$, are shown in the two panels
of Fig.~\ref{fig:comparo3vs}.  As in Figs.~\ref{fig:radialcomparo} and
\ref{fig:orbitalcomparo}, the scale of length is set by $x_0$.
All three trajectories have an acceleration $a_{\rm cross}=0.0005/x_0$.
As a rough estimate, this acceleration multiplied by the damping
time ($2\pi$ divided by the imaginary part of the TDP QN 
frequency $1/2x_0$) gives approximately $\Delta v=0.01$, on the order of a tenth 
of the crossing speed of the trajectories. It should, therefore,
be a reasonable approximation to consider the infall speeds to be constant 
during the plunge and excitation of QNR. (Values of acceleration much 
smaller than $0.0005/x_0$ turn out not to be as useful in demonstrating
features of the results.)

In the left panel the initial positive peaks are {\em not} QNR. Those
large peaks, prior to QNR, are analogous to the GWs from the pre-QNR
plunge, and we shall call these the ``plunge peaks.'' The QNR, a much
less prominent feature of the profile, starts around $u=u_{\rm
  cross}$, and can be seen as a small negative dip. The stark contrast
of this subtle feature with the very visible QNR in the Schwarzschild
problem is ultimately due to the fact the QNR damping frequency is
equal to the QN oscillation frequency in the case of the TDP. In the
Schwarzschild case, the QNR damping frequency is about a quarter of
the QNR oscillation frequency.

What is particularly important to notice is that the QNR in the right
panel does not agree with the predictions in Eqs.~(\ref{eq:upeak}) and
(\ref{eq:psipeak}). The QNR peak does not occur reasonably close to
$u-u_{\rm cross}=\pi/4$, nor is the peak value of $\Psi/x_0$
approximately -0.19 as
predicted by (\ref{eq:psipeak}). The dependence on $v_{\rm cross}$, in
fact is markedly different from that in (\ref{eq:psipeak}); while the
equation predicts a peak that decreases very slightly with increasing
$v_{\rm cross}$, the results show a significant {\em increase} in the
peak with increasing $v_{\rm cross}$.  This cannot be attributed to
the fact that the infall speed is not exactly constant; in fact, the numerical evaluation of only the
``late'' integral for the trajectories in Fig.~\ref{fig:comparo3vs}
agrees to high accuracy with the predictions of Eqs.~(\ref{eq:upeak})
and (\ref{eq:psipeak}). Rather, 
we learn from Fig.~\ref{fig:comparo3vs} that the QNR is dominated by 
the ``early'' integral, the second integral in
Eq.~(\ref{eq:simpleint}). This is somewhat surprising, and  is the first
indication of how one might think about the origin of what is the
dominant part of GW radiation in a binary merger.

\begin{figure} 
  \includegraphics[width=0.48\textwidth]{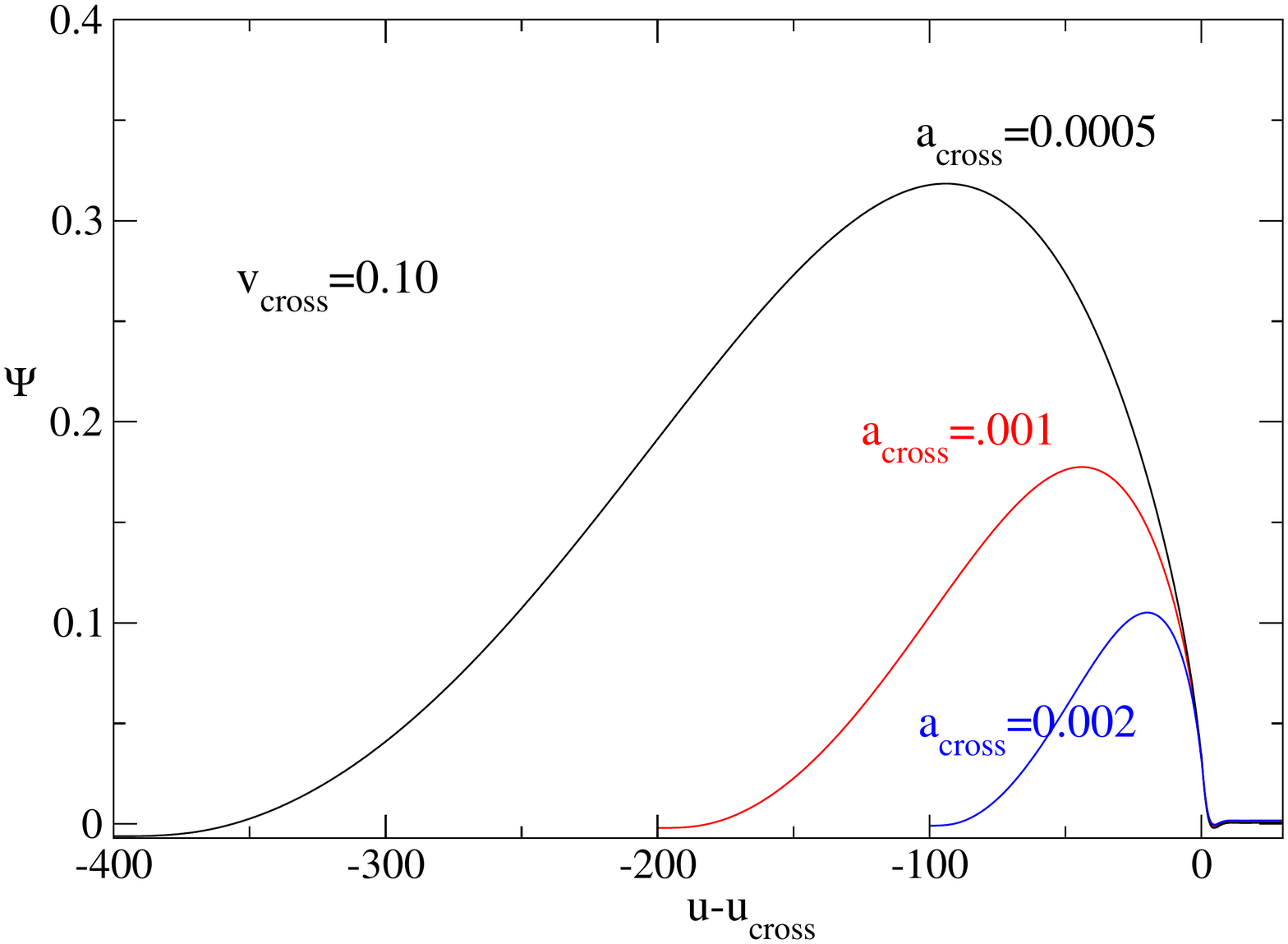}
\includegraphics[width=0.48\textwidth]{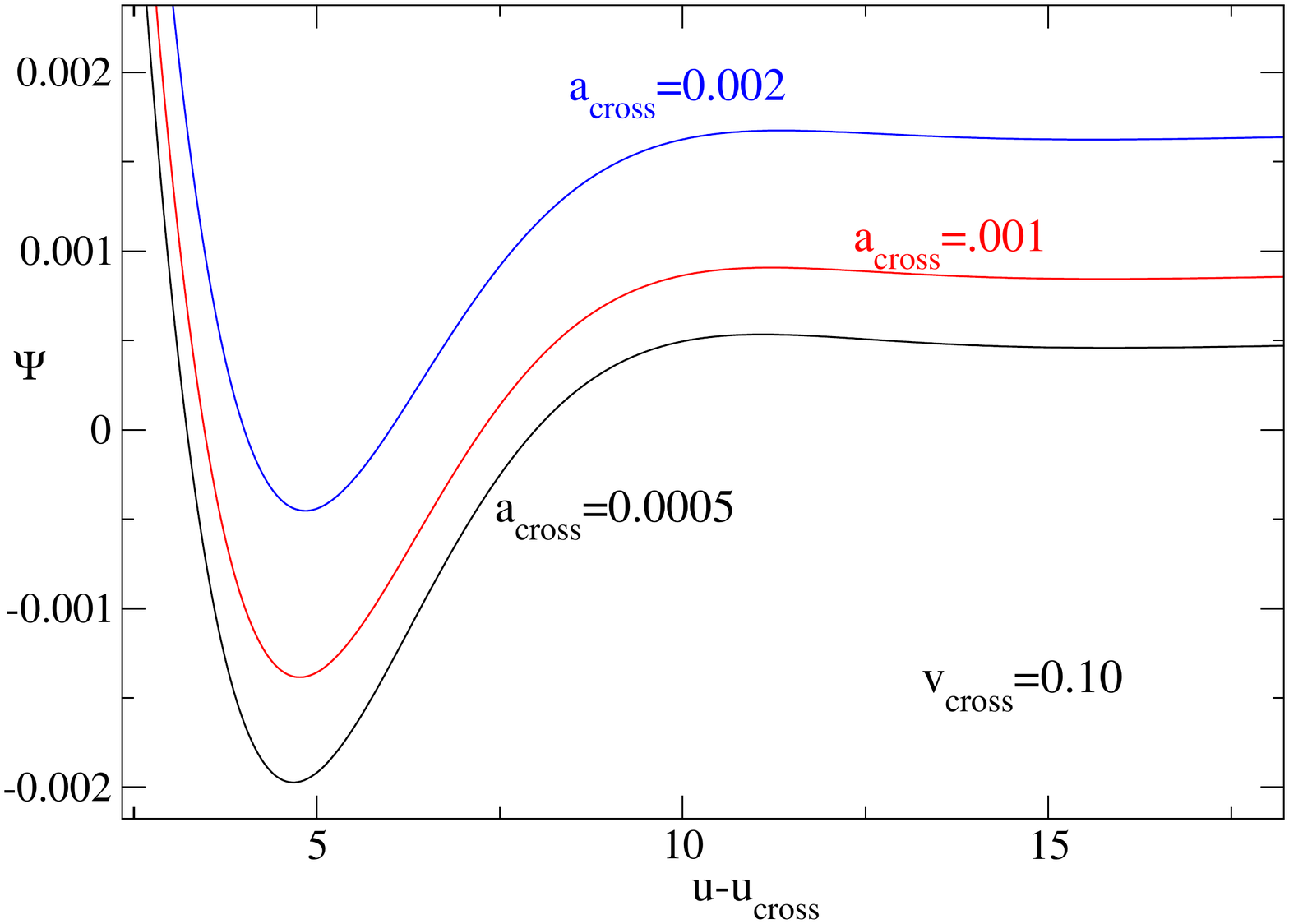}
\caption{The radiation due to a particle in radial infall in the TDP
  background. The horizontal axis is the retarded time \(u =
  t-x\). (See the text for an explanation of units.) The retarded time
  when the first signal from the potential edge arrives is indicated
  by \(T_{\textrm{cross}}\). Also shown in the figure is the
  quasinormal oscillation, the only half cycle discernible at this
  scale for the velocity and acceleration at crossing.}
\label{fig:comparo3as}
\end{figure}

The question remains whether the plunge and QNR are determined only by
the infall speed.  In flat spacetime, after all, the generation of
radiation is associated with acceleration. There is no hint of this
in the integrals in Eq.~(\ref{eq:simpleint}), and it is confirmed with
the numerical results in Fig.~\ref{fig:comparo3as} for different
accelerations. With an acceleration of $0.001/x_0$, for a damping
time $4\pi x_0$, there should be an increase in infall speed on the
order of $\Delta v\sim0.01$.  The increase in acceleration, with
$v_{\rm cross}$ fixed implies a larger speed after crossing and a
smaller speed before. In Fig.~\ref{fig:comparo3vs}, we saw that 
a decrease in both the plunge peak and the QNR excitation is associated
with a smaller speed (in the case of a very nearly constant speed).
The decrease in plunge peak and QNR excitation therefore reinforces
the conclusion that it is the particle motion prior to edge crossing
that plays the dominant role in generating this radiation. The argument
is strengthened by noting that to very rough order of magnitude the 
effect of the change of acceleration is of the right size.

\section{Plunge and QNR for trajectories in the Schwarzschild geometry}\label{sec:Schw}

We emphasize that the simple TDP models above are exploited to give
insights into the way in which the trajectories influence the plunge radiation
and QNR. Whether they give the right insights for the black hole
problem is a distinct question, a question we take up in this section.
We use second-order Lax-Wendroff, finite-difference evolution
codes\cite{TeukCode} for the Teukolsky function $\psi_4$, with our
flexible cubic trajectories, to investigate whether, as in the models
of the previous section, the plunge and QNR excitation depend only on
the conditions at the ``edge'' of the potential. For this purpose, we
take the edge of the potential to be at $r=3M$.  (The physical
justification for the importance of the LR is discussed in
Sec.~\ref{sec:discussion}.)

Some comments need to be made about the way in which we judge and
compare the size of QNR.  Two issues arise here: First is the fact
that unlike the ringing of normal modes, QNR has an amplitude that
depends on time. We must therefore find a way in which the magnitude
we ascribe to QNR is independent of the time at which we measure the
amplitude. The second issue is that we are only looking at the real
part of $\psi_4$; might the imaginary part contain different
information?

A satisfactory solution to both of these issues is to ``remove'' the
damping from the radiation.  To compare results for models with
different trajectories, we can multiply all curves to be compared by
$\exp{(0.088965\,t/M)}$, using the imaginary part, $0.0889/M$, of the
dominant (least damped) quadrupole Schwarzschild QN mode, the mode of
greatest astrophysical relevance. This leads to QNR having a
time-independent amplitude, thereby facilitating a comparison of the
excitation of QNR with different phases. It also shows us that the
real and imaginary parts of $\psi_4$ have the same amplitude once the
damping is removed, so no information is lost by plotting only the
real part of $\psi_4$.

While this technique may be a very useful tool for working with QNR in
other contexts, here we will use it explicitly only near the end of this
section, to aid in comparing QNR with different phases. For the most
part, theœ main value of this technique in the current paper is to
assure us that we are not missing anything. The lessons contained in
the curves presented below demonstrate the insights of interest,
without a need for a more careful extraction of QNR amplitude.

\subsection{Models for radial infall}
\begin{figure}[h]
  \begin{center}
  \includegraphics[width=.5\textwidth ]{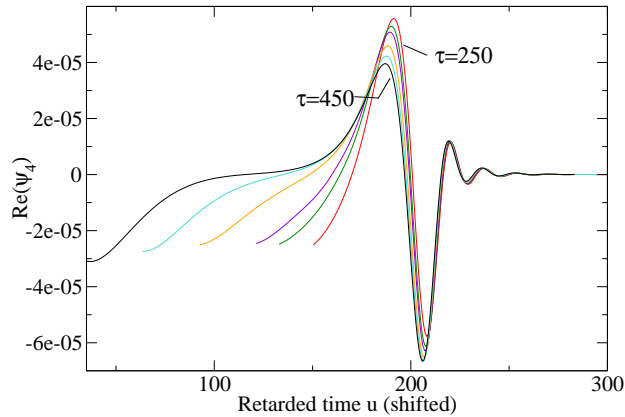}
  \caption{The plunge radiation and QNR in the real part of the Teukolsky function $\psi_4$
    for radial cubic trajectories, all with $v_{\rm LR}=0.3$ at the
    $r=3M$ LR, but with different values of the $\tau$
    parameter for the cubic trajectories of Eq.~(\ref{eq:FofT}).  The
    values of time are shifted so that all trajectories arrive at the
    $r=3M$ LR at the same time.  The values of $\tau$ range
    from 250 to 450. The peak around $u=190$ has a monotonically
    decreasing height with increasing $\tau$.  Note that this peak is
    not pure QNR, but rather a transition to QNR. The curves shortly
    after the peak show the QNR, and demonstrate remarkable agreement
    in the amplitude and phase of the QNR for the different
    trajectories. }
  \label{fig:Schw_difftauv3}
  \end{center}
  \end{figure}

\begin{figure}[h]
  \begin{center}
  \includegraphics[width=.5\textwidth ]{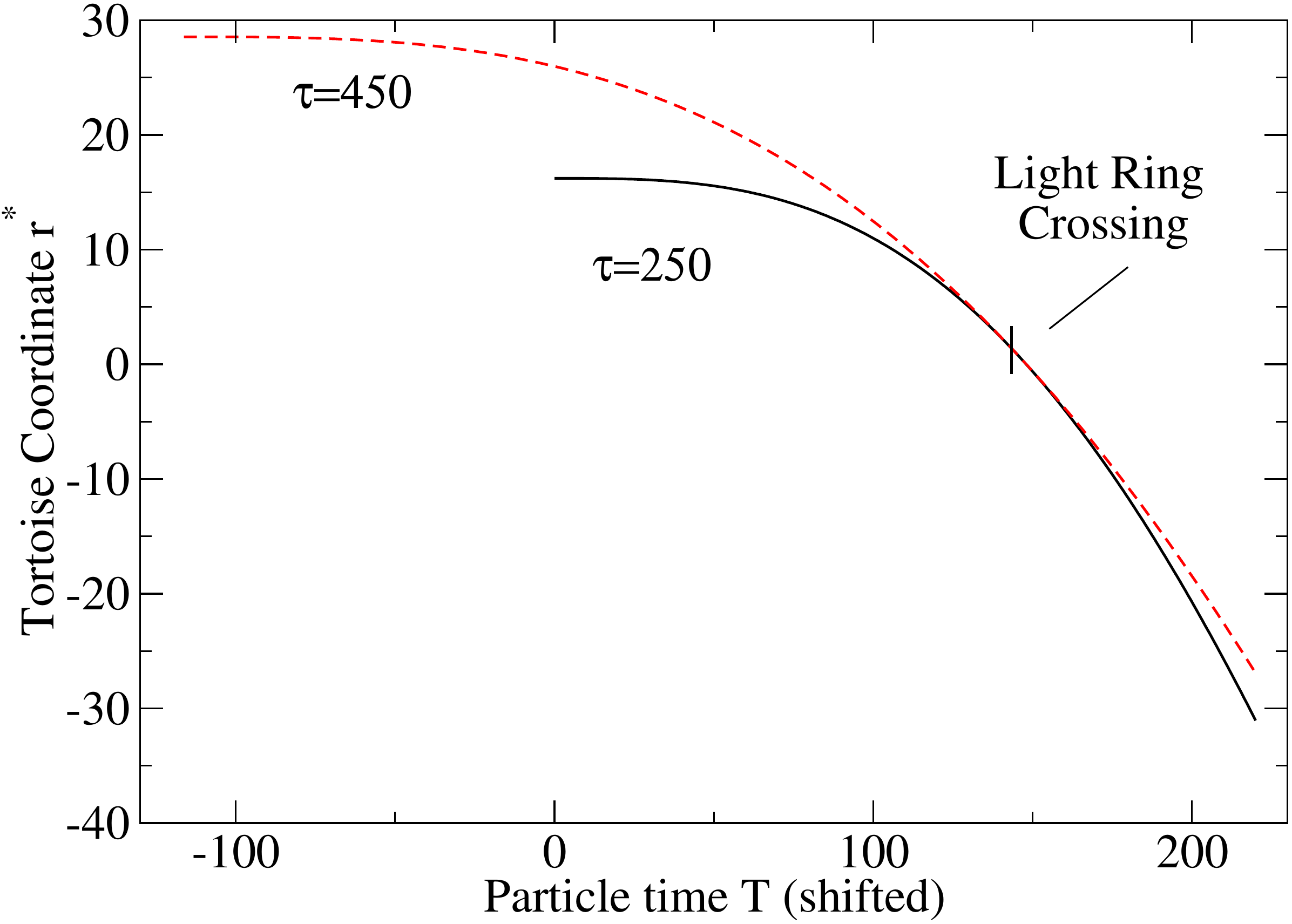}
  \caption{ Trajectories are shown for the $\tau=250$ and $\tau=450$
    cases of Fig.~\ref{fig:Schw_difftauv3}. The tortoise coordinate
    location $r^*$ of the particle is shown as a funtion of particle
    time $T$, according to the cubic trajectory of
    Eq.~(\ref{eq:FofT}).  For the $\tau=450$ trajectory, the time has
    been shifted so that the two trajectories arrive at the $r=3M$
    LR at the same value of $T$. Note that the trajectories
    differ significantly both before and after the LR
    crossing.}
  \label{fig:Schw_trajs}
  \end{center}
  \end{figure}

\begin{figure}[h]
  \begin{center}
  \includegraphics[width=.5\textwidth ]{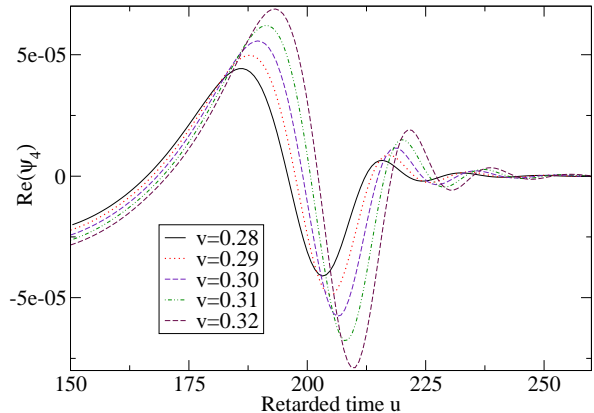}
  \caption{The radiation from the plunge and the QNR in the real part
    of the Teukolsky function $\psi_4$ for radial cubic trajectories,
    all with $\tau=250$, but with different values of $v=|dF/dT|$ at
    the LR ($r=3M$). The retarded time for the curves has been
    shifted, so that all trajectories pass the LR at the same
    time.  }
  \label{fig:Schw_diffv}
  \end{center}
  \end{figure}

We start by considering radial infall, with $r^*=F(t)$, in which $F$
is given by Eq.~(\ref{eq:FofT}), and in which $r^*$ is the solution of
Eq.~(\ref{eq:rstardef}) taken to be $r+2M\ln{(r/2M-1})$.  The models
that are the basis for 
Fig.~\ref{fig:Schw_difftauv3} 
are chosen all to have the same inward speed, $v_{\rm LR}=0.3$ at
the LR. (In this figure, and all those for the Schwarzschild
spacetime, numerical values are given in terms of the geometrized
($c=G=1$) Schwarzschild mass $M$.)  Different trajectories, with this
same LR speed, are created by varying the $\tau$ parameter of
Eq.~(\ref{eq:FofT}).  (The value of $a_0$ is then adjusted so that
$v_{\rm LR}=0.3$.) For a wide range of $\tau$ values,
Fig.~\ref{fig:Schw_difftauv3} shows that there is almost no difference
in the amplitude or phase of the QNR, and that there are  moderate
differences in the peak of the radiation at the plunge transition.
Though the QNR has negligible variation, the trajectories themselves
vary considerably.
We illustrate this in Fig.~\ref{fig:Schw_trajs}
which compares $r^*(T)$ for the extreme cases $\tau=250$ and $\tau=450$.
We show in Fig.~\ref{fig:Schw_diffv}
the opposite of what is shown in Fig.~\ref{fig:Schw_difftauv3}. Here
all trajectories have $\tau=250$, while the LR speed $v_{\rm
  LR}$ varies from 0.28 to 0.32.
From the form of the trajectories, we find that the acceleration at
LR crossing has a magnitude of approximately $2T_{\rm
  cross}/\tau^2$, which gives a value of approximately $0.0046/M$ for
$\tau=250$ and $0.0026/M$ for $\tau=450$.

  \begin{figure}[h]
  \begin{center}
  \includegraphics[width=.5\textwidth ]{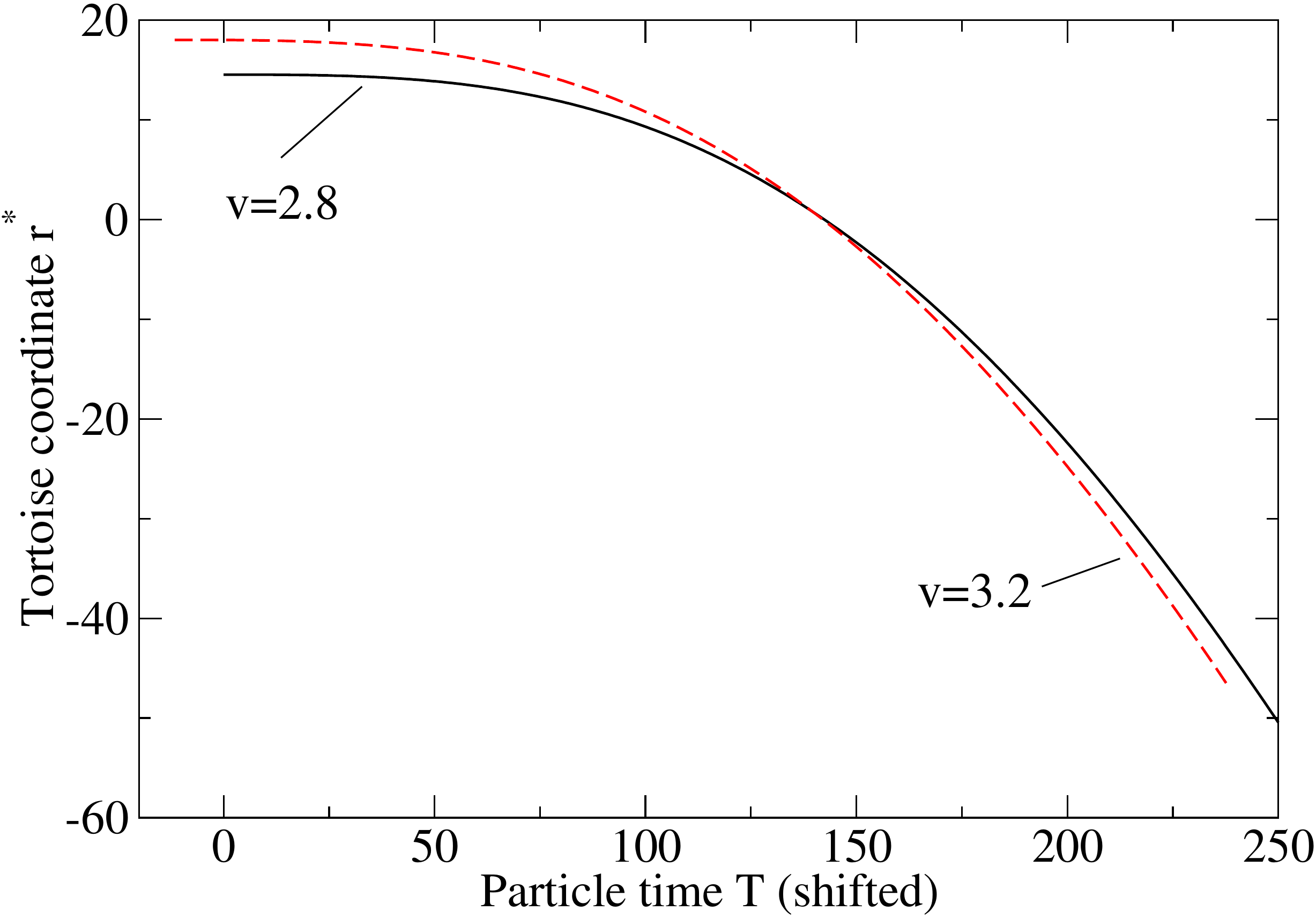}
  \caption{Trajectories for $v_{\rm LR}=0.28, 0.32$ and $\tau=280$,
    the two extreme examples in Fig.~\ref{fig:Schw_diffv}. The
    trajectories are exactly the same shape since in
    Eq.~(\ref{eq:FofT}) they differ only in $a_0$, and hence in the speed 
when they cross the LR. In the figure, the
    $v_{\rm LR}=0.32$ curve is shifted so that its LR crossing
    time agrees with that for $v_{\rm LR}=0.28$.  }
  \label{fig:trajstaufixed}
  \end{center}
  \end{figure}

What insights can be gained from a comparison of the TDP results and
the Schwarzschild results for radial infall?  In the Schwarzschild
case the LR plays the role of the edge. This ``edge'' is not sharply
defined as in the TDP case. Another, perhaps more important difference
from the TDP case is the longer damping time. These differences result
in an important difference in the nature of the generation of the
plunge radiation and QNR.  In the TDP case this generation was
dominated by the early, pre-edge, motion. In the Schwarzschild case
the generation seems to be dominated by the late motion. Notice in
particular that in Fig.~\ref{fig:Schw_difftauv3} the retarded time
(after crossing the LR) and amplitude of the QNR are almost
independent of the speed at the LR. This was a specific prediction for
the TDP provided that the QNR is dominated by the particle's motion
after crossing the potential edge. This prediction translates very
well to the Schwarzschild case since the late (post-LR) contribution
should be described by a FDGF like that in Eq.~(\ref{eq:TDPFDGF}),
with the TDP poles replaced by the Schwarzshild quadrupole GW poles.
(To be sure, there will be additional functions of $\omega$, but these
will play the role of constants in the evaluation of the residues at
the QN poles.)  This implies that the late QNR excitation will be
governed by an integral analogous to the late integral in
Eq.~(\ref{eq:simpleint}). The results in Fig.~\ref{fig:Schw_difftauv3}
therefore tell us that the late radiation dominates the excitation of
QNR.

Another, somewhat surprising, conclusion can be drawn from these
results for Schwarzschild radial infall.  Figure~\ref{fig:Schw_diffv}
shows the plunge radiation increasing with increasing speed at the
LR. This is analagous to the TDP plunge radiation increasing
with increasing speed at the potential edge.  Insights can then be
taken from Fig.~\ref{fig:Schw_difftauv3}. The plunge radiation is
larger for larger acceleration (smaller $\tau$) at a given crossing
speed. This is the reverse of what is seen in the TDP results, for which
greater acceleration means smaller plunge radiation. Just as the 
TDP results indicated that the plunge radiation is dominated by 
the early (pre-crossing) motion, the Schwarzschild results indicate
that the plunge radiation is dominated by the late, post-crossing radiation.

\subsection{Models for orbital trajectories}
As in the radial case, it should only be the motion near the LR that
is relevant to the generation of plunge radiation and the excitation of QNR. This suggests that when there is
angular velocity $\omega\equiv d\phi/dt$, it is the angular velocity
$\omega_{\rm LR}$ at the LR that is important, as well as the radial
velocity $v_{\rm LR}$.

If we are to investigate our
hypothesis that plunge radiation and  QNR excitation is determined only by the
conditions, in this case the angular velocity, at the LR, then we need
at least a two-parameter family of trajectories for $\omega(r)$ or
$\omega(t)$, and for clarity of analysis we would like to vary the
orbital motion for a fixed choice of radial motion.
As we have explained for radial motion, the models need not be limited
to families of trajectories that are physically correct motions. In
the case of angular motion, however, we {\em do} need to consider
several physical limitations on the nature of $\omega$.  First, the
orbital angular velocity $d\phi/dt=\omega$ is redshifted. That is,
$\omega$ dies off as $(1-2M/r)$ as the particle approaches the
horizon.  It turns out that we must choose orbital motions that
respect this condition. If we do not, then the plunge radiation and
QNR, for interesting values of $\omega_{\rm LR}$, are hidden by the much
larger radiation coming directly from the particle motion.

A second feature of physically correct angular motion is that
$d\omega/dt=0$ at the LR. If we were to retain this feature it would
hamper the demonstration that it is only the value of $\omega$ at and
near the LR that is important. Though this feature hampers our
demonstration, it supports our hypothesis. Since $\omega$ near the LR
is not very different from at the LR, our hypothesis would be wrong
only if conditions on $\omega$ far from the LR would be important.

A third, and obvious, physical feature of orbits is that they must not
exceed light velocity. In our notation, with $v_{\rm LR}\equiv -
dr^*/dt$ at the LR, this condition is $27\omega^2_{\rm LR}M^2<1-v_{\rm
  LR}^2$. Though this is a physical constraint on $\omega_{\rm LR}$ it
need not be a constraint on our exploration of the nature of the
plunge radiation and QNR.

Our choice is to take the angular motion to be
\begin{equation}\label{eq:model}
\omega=\omega_{\rm LR}\,\frac{27M^2}{(1+\sigma)r^2}\left(1-\frac{2M}{r}\right)
\left(1+\frac{3\sigma M}{r}\right)\,,
\end{equation}
in which $\omLR$ appears explicitly as one of the parameters.  For all
values of the two parameters, $\omega$ in this model has the
physically correct redshift.  The parameter $\sigma$ governs
$d\omega/dt$ at the LR. For $\sigma>0$, in our models, the angular
frequency $\omega$ increases as the particle moves inward past the LR.

\begin{figure}[h]%
  \begin{center}
  \includegraphics[width=.5\textwidth ]{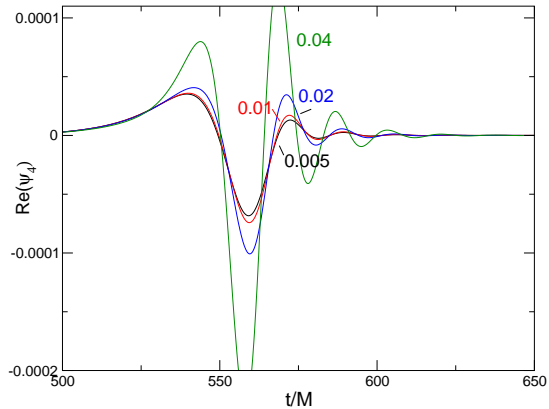}
  \caption{Plunge radiation and QNR excitation from models with
    angular frequency $M\omLR=$0.005, 0.01, 0.02, 0.04.  All models
    have radial motion for $a_0=38.6114\,M$ and $\tau=600\,M$, and
    hence in all cases the particles pass the LR at time
    $T_\times=348.90\,M$, with a speed $v_{\rm LR}=0.3$. For all
    orbital motions, $\sigma=1.$ Note that each waveform shown is
    extracted at $r^*=200\,M$, implying that the waveform at
    $T\approx550$ originates at the LR crossing.  }
  \label{fig:versmall}
  \end{center}
  \end{figure}

\begin{figure}[h]
  \begin{center}
  \includegraphics[width=.5\textwidth ]{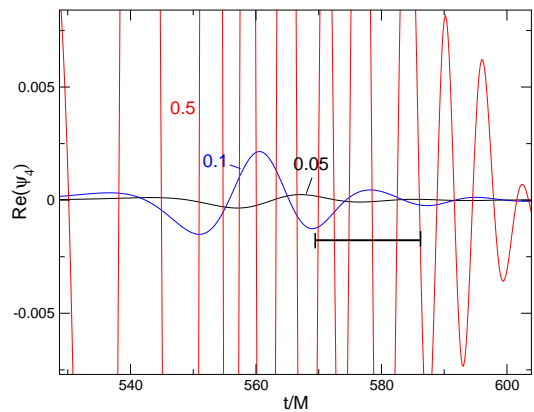}
  \caption{Plunge and post-plunge radiation for models with
    $M\omLR=$0.05, 0.10, and 0.50. The radial fall is that of
    Fig.~\ref{fig:versmall}.  The dark horizontal bar shows the period
    $\Delta t=16.815\,M$ for a full wave of QNR.  For $M\omLR=0.50$
    the more rapid oscillations correspond approximately to the
    orbital frequency, not to the QN frequency.  }
  \label{fig:small}
  \end{center}
  \end{figure}

We start the investigation of the effects of orbital motion by asking
what is the interesting range of values of $\omLR$.  Results of
evolution codes are shown in Figs.~\ref{fig:versmall} and
\ref{fig:small}. For the models shown the radial speed as the LR is
crossed is $v_{\rm LR}=0.3$, so that the speed-of-light limit on
$\omLR$ is $0.1836/M$. And for all models shown in
Figs.~\ref{fig:versmall} and \ref{fig:small}, the $\sigma$ parameter
in Eq.~(\ref{eq:model}) is taken to be $\sigma=1$.  The results in
Fig.~\ref{fig:versmall} are for very small values of $\omega_{\rm
  LR}$, values that do not exceed the light speed limit. For all these
models, the post-plunge radiation (waveforms after $t/M$ is 548.9) is
QNR. For the smallest value $\omLR=0.005/M$ the result is
indistinguishable from pure radial infall. This is to be expected
since this value is much smaller than the imaginary (damping) part of
the QNR frequency, $0.088965/M$. From the FDGF point of view this means
that the motion of the source due to angular velocity is insignificant
until the damping factor has reduced the QNR excitation integral to a
negligible size.

As $\omLR$ increases to a value
comparable to the imaginary part (0.0889/M) of the QNR frequency, the
orbital motion starts to have an effect, increasing both the plunge
radiation and the excitation of QNR. For $\omLR=0.01/M$ both can be
seen to be distinguishable from that for $\omega_{\rm
  LR}=0.005/M$. With a further increase of the value of $\omLR$, to a
value still within the speed-of-light limitation, the effect of
orbital motion becomes dramatic.

Figure \ref{fig:small} shows the post-plunge radiation for values of
$\omLR$ approaching and exceeding the light speed limit. In the models
shown, the post-plunge radiation for both $\omLR=0.05/M$ and $0.10/M$
have a period of the same order as the period 16.81\,$M$ of the least
damped Schwarzschild QN mode. (This period is shown as a horizontal
bar.) For $\omLR=0.5/M$, the very large oscillations are at a clearly
different and noncostant frequency. As models progress from
$\omLR=0.1/M$ to higher frequency, the late time results are no longer
pure QNR, but are QNR increasingly mixed with some higher frequency
radiation.  Below, in subsection~\ref{subsec:direct}, we demonstrate
that for $\omLR$ well above the QNR frequency what we are seeing is
radiation coming directly from the motion of the particle.

\begin{figure}[h]%
  \begin{center}
  \includegraphics[width=.5\textwidth ]{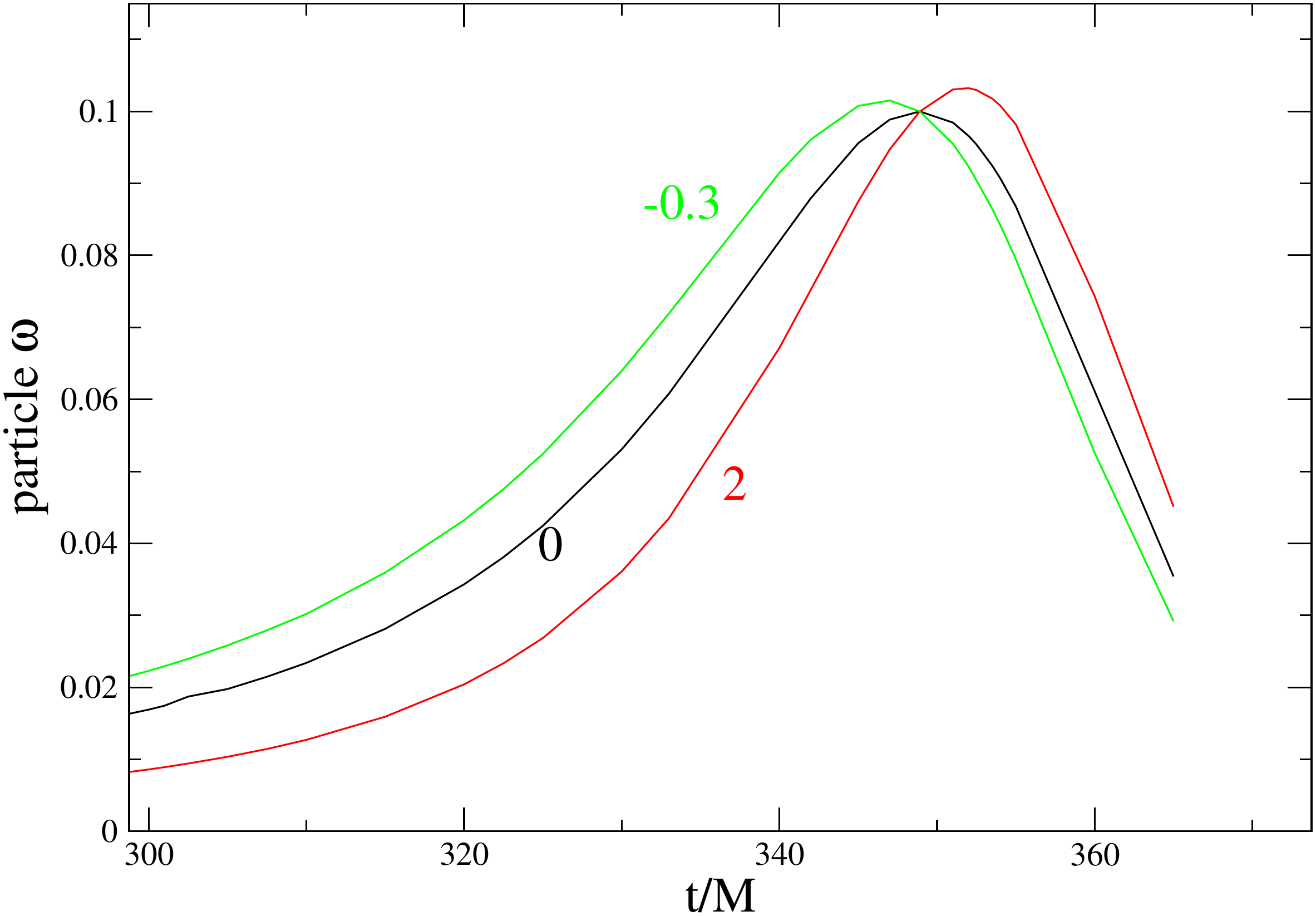}
  \caption{Particle angular velocity as a function of time for models
    following Eq.~(\ref{eq:model}) with $\omLR=0.1$, but with
    $\sigma=0, 2, -0.3$. The $\sigma$ values are indicated in the
    plot.}
     \label{fig:om4diffsig}
  \end{center}
  \end{figure}

Orbital frequencies well above that of QNR are unphysical, so we
presently limit our considerations to $\omLR$ below 0.10 or so, and we
return to the central question of this paper in the context of this
subsection: Are plunge radiation and the excitation of QNR determined
primarily by $\omLR$, and not by the angular motion at other radii? To
investigate this question, we consider a family of trajectories, all
with the radial motion used throughout this subsection ($\tau=600$,
$v_{\rm LR}=0.3$) and all with $M\omLR=0.1$, a physically reasonable
choice for which QNR is not dominated by direct radiation.  We choose
the models of Eq.~(\ref{eq:model}) with $\sigma=0,\, 2,$ and -0.3.
For these choices Fig.~\ref{fig:om4diffsig} shows that the time
dependence of the orbital frequency differs significantly. The
sensitivity to changes in the particle's orbital velocity can be
inferred from the data in Fig.~\ref{fig:small}, where a change in
$M\omLR$ from 0.05 to 0.1 results in a sixfold increase in the
amplitude of QNR. By that standard, these models should have quite
different excitation of QNR if the particle angular velocity far from
the LR were significant.

\begin{figure}[h]%
  \begin{center}
  \includegraphics[width=.5\textwidth ]{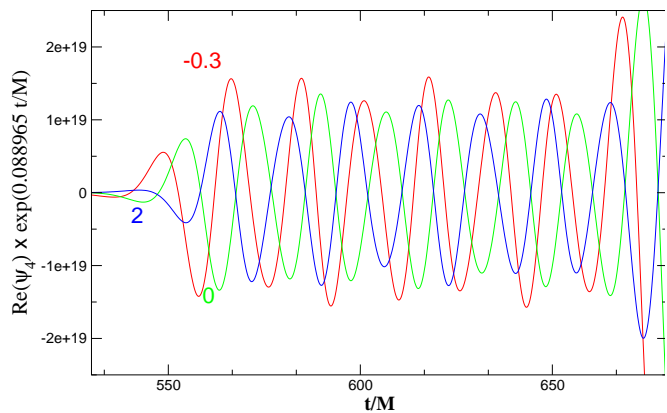}
  \caption{The plunge radiation and QNR (real part of the Teukolsky function)
    for the infall models of this subsection. All curves here
    correspond to $\omLR=0.10$, but their orbital motion, as
    prescribed in Eq.~(\ref{eq:model}), has different values of the
    $\sigma$ parameter, the values $\sigma=-0.3$, 0, and 2, as marked
    on the plot. Each curve has been multiplied by a factor
    $\exp{(0.088965\;t/M)}$ to remove the QN exponential decay. This
    greatly clarifies when QNR starts as well as facilitates the
    comparison of QNR with different phases.}
   \label{fig:diffsigmas}
  \end{center}
  \end{figure}

Figure~\ref{fig:diffsigmas} shows the computed plunge radiation (real
part of the Teukolsky function) for the three models. Comparison of
the amplitude of QNR cannot be made directly from plots of $\psi$ due
to the phase differences in the radiation for the three models. We
therefore apply the technique discussed at the beginning of this
section, the technique of ``removing'' the damping by multiplying by
$\exp{(0.088965\;t/M)}$. The ``undamped QNR,'' plotted in
Fig.~\ref{fig:diffsigmas}, shows the QNR as a series of oscillations
of approximately constant amplitude. The appearance of this feature
adds confidence that the radiation, during this epoch, is QNR, and it
greatly narrows the uncertainty about where the QNR starts. Most
important for our current concern, it also allows a direct comparison
of the amplitudes for the different models.

From these results, with damping removed, we find that the QNR
amplitudes for the $\sigma=-0.3$ model is roughly 15\% greater than that
of the $\sigma=2$ or $\sigma=0$ models.  Some variation, of course,
should be expected since it is not precisely the $\omega$ {\em at} the
LR that is crucial, but rather the $\omega$ {\em near} the LR. Here
``near" means within a e-damping time $\Delta t=M/0.0889\approx
11M$. The variation in the models over that time makes it surprising
that there is as little variation in QNR excitation as is computed;
the principle that the LR motion is crucial is therefore strongly
supported.  The conclusion is yet stronger for physically correct
models of geodesic motion, since those models have $d\omega/dt=0$ at
the LR.

\subsection{Direct radiation}\label{subsec:direct}
Here we demonstrate that for orbital frequencies $\omLR$ well above
the frequency of Schwarzschild QNR, $M\omega=0.37367$, the radiation
is dominated by a field directly linked to the motion of the
particle. With this we can infer that for frequencies near and
somewhat larger than $M\omega=0.37367$, the radiation will be a
mixture of such direct radiation and QNR. We choose the specific case
$M\omLR=4.0$, an order of magnitude above the frequency of QNR, a
frequency at which we should see the direct radiation completely
dominate the QNR.

\begin{figure}[h]%
  \begin{center}
  \includegraphics[width=.6\textwidth ]{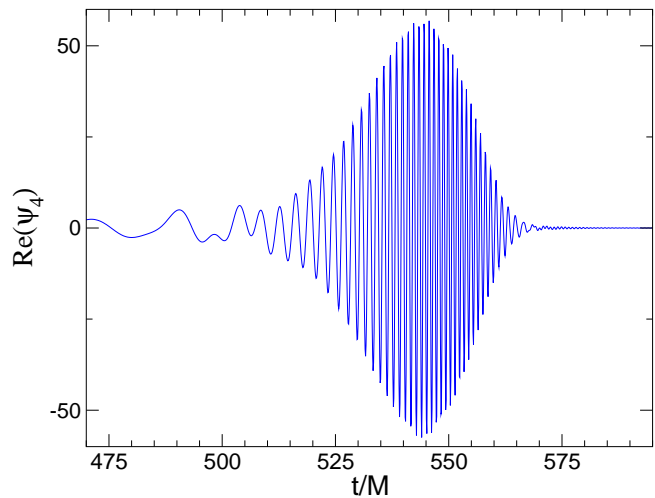}
  \caption{The real part of the Teukolsky function for our infall
    model with $\omLR=4$. The $\ell,m=2,2$ part of the solution is
    presented. }
  \label{fig:projected}
  \end{center}
  \end{figure}

To analyze the radiation at this high frequency we first need to deal
with the fact that the orbital motion is so fast that many angular
modes of radiation are generated and we lose the simplicity of a
single multipole moment.  For that reason, we present in
Fig.~\ref{fig:projected} only the $\ell,m=2,2$ part of the radiation.

Several features of Fig.~\ref{fig:projected} are immediately
interesting.  The peak of the emission is around $t/M\approx547$, the
time at which the retarded location of the particle orbit is at the
$r=3M$ location of the LR.  The ``edge'' of the curvature
potential in the Schwarzschild spacetime is, of course, not a sharp
well defined edge as in the TDP, so the LR is only a nominal
representation of an edge location. A second point to notice is that
the increasing amplitude for $t/M\lesssim547$ can be ascribed to the
increasing particle acceleration, and the decreasing amplitude for
$t/M\gtrsim547$, can be understood as a redshift effect.

\begin{figure}[h]%
  \begin{center}                        
  \includegraphics[width=.4\textwidth ]{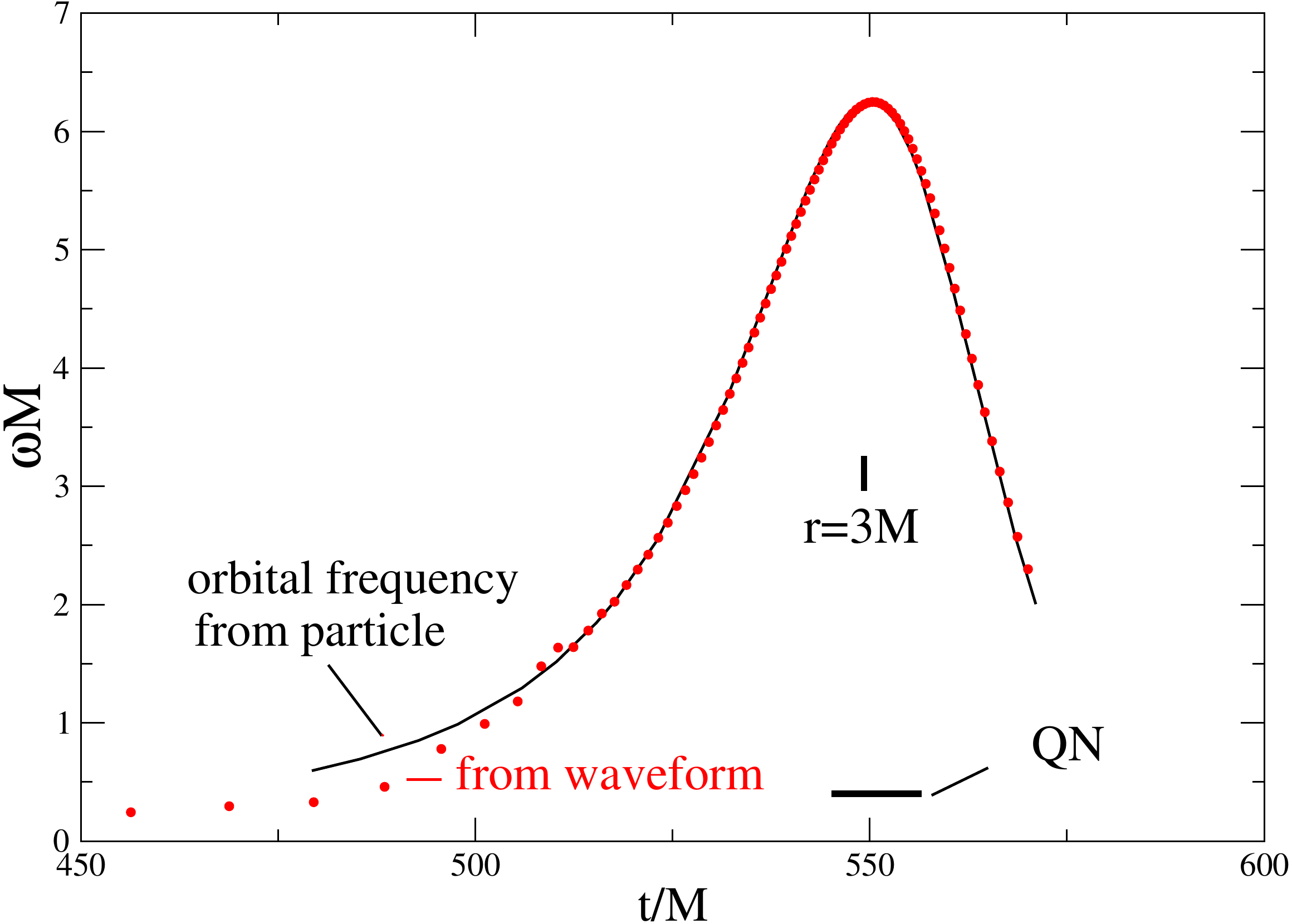}
  \caption{ A comparison of the frequency seen in the evolved waveform
    (circles) and the frequency inferred from the particle motion
    (solid curve). The waveform results above $t\approx570$ are noisy
    due to the small amplitude of the computed waves. Five-point
    averaging was applied to smooth the waveform results below
    $t\approx515$. The horizontal bar shows the value of a period
    corresponding to the frequency $\omega M=0.37367$
    for the least damped QN quadrupole mode.  See text for details of
    the computation of the frequency inferred from the particle
    motion.}
    \label{fig:comparo}
  \end{center}
  \end{figure}

To delve more deeply into the nature of the ``direct'' radiation in
Fig.~\ref{fig:projected} we plot, in Fig.~\ref{fig:comparo}, a
comparison of the frequency exhibited by the computed wave (inferred
from the spacing between the zeros of the waveform) and the frequency
that one expects for radiation coming directly from the particle's
orbital motion. This expected frequency is computed by noting, for a
particular ``emission time'' in the particle's trajectory, the angular
frequency of the particle as given by Eq.~(\ref{eq:model}). The
``observation time'' is then computed for the radiation from this
event to reach the wave extraction radius $r^*=200M$; it is this
observation time that is the horizontal axis in Fig.~\ref{fig:comparo}. The
frequency is then multiplied by two to account for the fact that the
gravitational waves are $m=2$. Lastly, the Doppler shift of the
outgoing radiation is taken into account by dividing the frequency by
$1+v$, where $v$ is the ingoing radial speed of the particle at the
emission time. The excellent agreement between this prediction and the
waveform observed leaves no doubt that this is ``direct'' radiation,
and unrelated to the relatively low quasinormal frequency (indicated
in the plot by a short horizontal line).

The analysis of direct radiation may turn out to be more than a useful
check on consistency; it may turn out to be an interesting diagnostic
in the Kerr spacetime. In the Schwarzschild spacetime, the amplitude
vs.\ frequency result has, embedded in it,
information about the way in which the curvature potential acts as a
high pass filter for radiation passing outward.  It has been
suggested\cite{AlessandraB} that amplitude vs.\ frequency curves may
lead to an effective Kerr curvature potential.

\section{Discussion}\label{sec:discussion}
Though great attention has been given to the connection of QN
frequencies and the
LR\cite{Goebel,PoschlTeller,BuCoPr:2006,Berti:2007,Zim:2012,Dolan:2011},
much less has been given to the question of why the LR should play
such a crucial role in the generation of plunge radiation and 
the {\em excitation} of QNR?  From the point of view of multipole
decomposition and the curvature potential, this is because the
curvature potential takes a simple flat spacetime multipole form for
$r\gg3M$, and is negligibly small far on the other side of the LR
(more precisely for $1-2M/r\ll1$).  But this observation is not an
explanation. An explanation must be rooted in the physical process
being described before multipoles are projected.

That physical process is the generation of disturbances by the motion
of a particle.  Disturbances propagate along characteristics, the same
world lines as the paths of photons. Consider a particle and an
observer both far outside ($r\gg3M$) a black hole. Disturbances
(scalar or whatever) of the particle will travel on the usual
characteristics connecting the moving particle to the observer. Due to
the presence of the black hole, however, there will be extra,
non-flat-spacetime characteristics connecting the particle and
observer. These will correspond to zoom-whirl orbits that go around
the black hole, just outside the LR, and end up at the
particle. Though interesting as a matter of principle, these extra
``nonstandard'' characteristics will play little role if the black
hole is far away; the cross section for such characteristics will be
too small.

When the particle is close to the LR, however, the effect of the
zoom-whirl characteristics will be dramatic. These ``extra''
characteristics will dominate over the standard characteristics as the
particle passes through the LR.  It is expected, and confirmed by
studies of trajectories, that as the particle proceeds inward, its
characteristics more and more lose the ability to reach the external
observer and the particle and observer become more and more causally
disconnected.
This picture of the zoom-whirl characteristics give a fairly
persuasive heuristic argument that radiation that will be
characteristic of the black hole, plunge radiation and QNR, must come
from the region of the LR.

\section{Conclusion}\label{sec:conc}
In this paper we have used the Fourier domain Green function to
understand why the excitation of plunge radiation and quasinormal
ringing might be determined by source conditions at the light ring of a
black hole.  We then verified that hypothesis of light ring cruciality with
models both using Green function computations and numerical evolution.
A particular insight informed by both methods was that the plunge
radiation and quasinormal ringing, at least in the case of radial
infall, can be primarily ascribed to the particle motion after 
crossing the light ring.

Such insights can be viewed as a  successful first step in
understanding the plunge/ringing phase of binary black hole
inspiral/merger, but only the first step. In particular, this work has
relied on the particle perturbation linearization of general
relativity. There are good reasons to believe that the lessons from
that simplification apply to the binary motion of comparable mass
holes, but this needs to be investigated.

It will be necessary to extend our study to rotating holes, an
extension which will be particularly interesting since there are two
distinct light rings in the equatorial plane, one for prograde photon
orbits and one for retrograde orbits, and there are light rings for a
continuum of radii if nonequatorial orbits are considered.  Work on
this analysis is already underway.

\section*{Acknowledgments}
We thank Scott Hughes of MIT and Alessandra Buonanno, of the Max
Planck Institute for Gravitational Physics, for very useful
discussions of several aspects of this work.  
We express sincere and profound thanks to two exceptionally 
conscientious referees whose careful reading of the original 
version of this paper have led to a much clearer and less error-prone  presentation.
SN acknowledges support
from the Center for Gravitational Wave Astronomy at UTB.
GK acknowledges research support from NSF Grant Nos. PHY-1303724, PHY-1414440
and from the U.S. Air Force agreement No. 10-RI-CRADA-09.

\appendix
\section{Fourier Domain Green Function for the TDP}\label{sec:AppA}

\subsection{Method}

We give here the details of the FDGF for the simple TDP model. The
method is to find a solution of Eq.~(\ref{eq:FDGF}) for the potential
in Eq.~(\ref{eq:TDPpot}).  The model is simple because the homogeneous
solutions of Eq.~(\ref{eq:FDGF}) are the elementary functions
\begin{equation}
  {\cal G}\propto\left\{
  \begin{array}{ll}
e^{i\omega x},\ e^{-i\omega x}\ &
{\rm for}\ x<x_0
\\e^{i\omega x}\left(1+i/(\omega x)\right),\ e^{-i\omega x}\left(1-i/(\omega x)\right)\                & {\rm for}\ x>x_0
  \end{array}
\right.    \,.
\end{equation}
A solution for ${\cal G}$ is constructed by imposing outgoing boundary
conditions (${\cal G}\propto e^{i\omega x}\left(1+i/\omega x\right)$)
for $x>{\rm max}(a,x_0)$, and ingoing boundary conditions (${\cal
  G}\propto e^{-i\omega x}$) for $x<{\rm min}(a,x_0)$. The values of
$x,x_0$ and $a$ then separate the solution into three forms with four
coefficients. The coefficients are determined by requiring that the
solution be continuous everywhere and have a unit discontinuity in its
derivative at $x=a$.

The form of the solution depends on the order of $x,x_0$ and $a$, and
the presentation of the solution is separated accordingly. We also present
the results for $\Psi$, using these FDGFs, for radial infall with $f(t)=1$.

\subsection{Solutions for $x>a>x_0$.}

\begin{figure}[h]
  \begin{center}
  \includegraphics[width=.25\textwidth ]{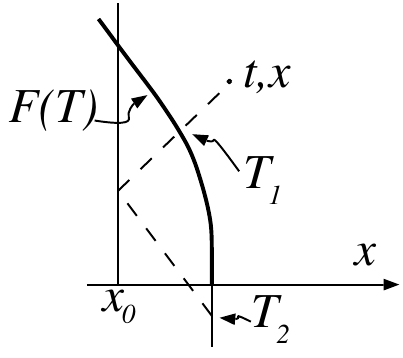}
  \caption{ A spacetime diagram showing the meaning of $T_1$ and
    $T_2$. The bold line represents the trajectory of the particle
    $x=F(T)$; the dashed lines are radial characteristics.  }
  \label{fig:T1T2}
  \end{center}
  \end{figure}

In this case the FDGF 
$$
 {\cal G}=\frac{e^{i\omega(x-a)}e^{-2 i \omega  {x_0}} \left((1+i a \omega ) e^{2 i \omega  {x_0}} \left(2 \omega ^2 {x_0}^2+2 i \omega  {x_0}-1\right)+e^{2 i a \omega } (1-i a \omega )\right)}{2 a \omega ^2 \left(2 \omega ^2
   {x_0}^2+2 i \omega  {x_0}-1\right)}
\left(1+\frac{i}{x\omega}\right)\,.
$$
A convenient rearrangement is
$$
  {\cal G}=
-\,e^{-i\omega(-x+a)}\left(\frac{1}{2i\omega}-\frac{1}{2a\omega^2}
+\frac{1}{2x\omega^2}
-\frac{i}{2ax\omega^3}
\right)\hspace{.9in}{\ }
$$
\begin{equation}\label{eq:convenient}
  -\,\frac{e^{-i\omega(-x-a+2x_0)}}{2x_0^2(\omega-\omega_1)(\omega-\omega_2)}
\left(
\frac{1}{2i\omega}+\frac{1}{2a\omega^2}
+\frac{1}{2x\omega^2}
+\frac{i}{2ax\omega^3}
\right)\,.
\end{equation}
Now we define the following two special particle times $T_1$ and $T_2$
as
\begin{equation}\label{eq:T1T2}
  T_1=t-x+F(T_1)\quad\quad T_2=t-x+2x_0-F(T_2)\,.
\end{equation}
The meaning of these two particle times is illustrated in the spacetime
cartoon in Fig.~\ref{fig:T1T2}.

With these definitions, and with Eq.~(\ref{eq:PsifromcalG}), we get
the following:
$$\Psi(t,x)=
-\,\frac{1}{2}\,\int_{-\infty}^{T_1}
\left[
1+\left(\frac{1}{x}-\frac{1}{a}  \right)\left(
t-T-x+a
\right)
-\frac{(t-T-x+a)^2}{2ax}
\right]\,dT
$$
$$
-\,\frac{1}{2}\,\int_{-\infty}^{T_2}
\left[1
+\left(\frac{1}{a}
+\frac{1}{x}\right)
(t-T-x-a)
+\frac{(t-T-x-a)^2
}{2ax}\right]\,dT
$$
\begin{equation}\label{psixgtra}
    -\,\frac{1}{2}\,\int_{-\infty}^{T_2} \,e^{-\xi}\left[
      -(\cos{\xi}+\sin{\xi})+2x_0\left(\frac{1}{a}+\frac{1}{x}\right)\cos{\xi}
      -\frac{2x_0^2}{ax}\left(\cos{\xi}-\sin{\xi}\right)
      \right]\,dT\,.
\end{equation}
Here $\xi$ is 
\begin{equation}\label{gammadef}
 \xi={(t-T-x-a+2x_0)}/{2x_0} \,.
\end{equation}

The first two integrals in Eq.~(\ref{psixgtra}) can be combined so
that the result is in manifestly finite form,
$$\Psi(t,x)=
-\,\frac{1}{2}\,\int_{T_2}^{T_1}
\left[
1+\left(\frac{1}{x}-\frac{1}{a}  \right)\left(
t-T-x+a
\right)
-\frac{(t-T-x+a)^2}{2ax}
\right]\,dT
$$
\begin{equation}\label{psixgtra2}
    -\,\frac{1}{2}\,\int_{-\infty}^{T_2} \,e^{-\xi}\left[
      -(\cos{\xi}+\sin{\xi})+2x_0\left(\frac{1}{a}+\frac{1}{x}\right)\cos{\xi}
      -\frac{2x_0^2}{ax}\left(\cos{\xi}-\sin{\xi}\right)
      \right]\,dT\,.
\end{equation}
Note that this solution is good only for $u\equiv t-x$ less than or equal to the
value of $u$ corresponding to $u_{\rm crit}=T_{\rm cross}-x_0$, where
$T_{\rm cross}$ is the time at which the particle reaches $x_0$.  For
larger values of $u$ the Green function for $a>x_0$ is no longer
valid.

\subsection{Solution for inner region $x<x_0<a$}

In this case the FDGF is 
\begin{equation}
  {\cal G}=\frac{e^{i\omega(a-x)}(1-ia\omega)}{2a(\omega-\omega_1)(\omega-\omega_2)}\,.  
\end{equation}
so that

$$ \Psi=\frac{1}{2\pi} \int_{-\infty}^\infty \int_{-\infty}^\infty
\frac{e^{-i\omega(t-T-a+x)}(1-ia\omega)}{2a(\omega-\omega_1)(\omega-\omega_2)}
\,dT d\omega $$

\begin{figure}[h]
  \begin{center}
  \includegraphics[width=.25\textwidth ]{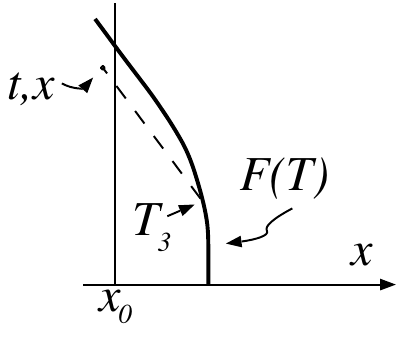}
  \caption{ A spacetime diagram showing the meaning of $T_3$. The bold
    line represents the trajectory of the particle $x=F(T)$; the
    dashed line is an ingoing radial characteristic.  }
  \label{fig:T3}
  \end{center}
  \end{figure}

We now define $T_3$ by $t+x=T_3+F(T_3)$, as illustrated in Fig.~\ref{fig:T3}, and we 
define 
$$
\sigma\equiv (t-T+x-a)/2x_0. 
$$
With these we get
\begin{equation}\label{eq:psiinner}
\Psi  = -\,\frac{x_0}{a}\int_{-\infty}^{T_3}
e^{-\sigma}\left[\left(1-\frac{a}{2x_0}\right)\sin{\sigma} 
+\frac{a}{2x_0}\cos{\sigma} \right]\,dT\,.
\end{equation}

\subsection{Solution for intermediate region $x_0<x<a$}

\begin{figure}[h]
  \begin{center}
  \includegraphics[width=.25\textwidth ]{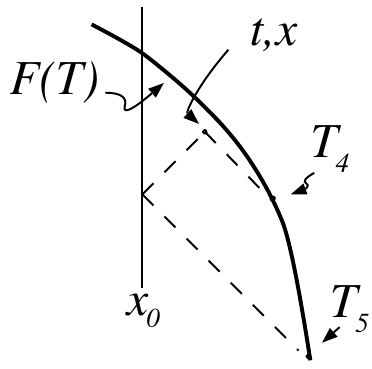}
  \caption{ A spacetime diagram showing the meaning of $T_4$ and
    $T_5$. The bold line represents the trajectory of the particle
    $x=F(T)$; the dashed lines are radial characteristics.  }
  \label{fig:T4T5}
  \end{center}
  \end{figure}

Here the FDGF is
\begin{equation}
{\cal G}=\frac{e^{-i\omega(x-a)}(1-ia\omega)}{2a\omega^2}\,\left(1-\frac{i}{\omega x}\right)  
-\,\frac{e^{-i\omega(-x-a+2x_0)}(1-ia\omega)}{4a\omega^2x_0^2(\omega-\omega_1)(\omega-\omega_2)}
\,\left(1+\frac{i}{\omega x}\right)\,.  
\end{equation}
We define $T_4$ and $T_5$ by
\begin{equation}
  T_4+F(T_4)=t+x\quad\quad\quad T_5+F(T_5)=t-x+2x_0\,.
\end{equation}
and illustrate them in Fig.~\ref{fig:T4T5}.

For evaluation of the integrals leading to $\Psi$ it is convenient 
to define
\begin{equation}
  K_4\equiv t-T+x-a\quad\quad\  K_5\equiv t-T-x-a+2x_0\quad\quad
\varepsilon\equiv K_5/2x_0\,.
\end{equation}
We now break up the integral for $\Psi$ into three parts 
and write $\Psi=I_1+I_2+I_3$, 
where 
\begin{equation}
  I_1=\frac{1}{2a}\int_{-\infty}^{T_4}\left[
-a-\left(1-\frac{a}{x}\right)K_4
+\frac{1}{2x}K_4^2
\right]\,dT
\end{equation}
\begin{equation}
  I_2=\frac{1}{2a}\int_{-\infty}^{T_5}\left[
-a+\left(1+\frac{a}{x}\right)
\left(-K_5+2x_0\right)
-\frac{1}{2x}\left(K_5^2-4x_0K_5+4x_0^2\right)   \right]\,dT
\end{equation}
\begin{equation}
  I_3=\frac{1}{2}\int_{-\infty}^{T_5}e^{-\varepsilon}\left[
    \cos{\varepsilon}+\sin{\varepsilon}
    -2x_0\left(\frac{1}{a}+\frac{1}{x}\right)\cos{\varepsilon}
    +\frac{2x_0^2}{ax}\left(\cos{\varepsilon}-\sin{\varepsilon}\right)
    \right]\,dT\,.
\end{equation}
We can combine $I_1$ and $I_2$ to give an answer that is manifestly finite
$$
\Psi= \frac{1}{2a}\int_{T_5}^{T_4}\left[
-a-\left(1-\frac{a}{x}\right)K_4
+\frac{1}{2x}K_4^2
\right]\,dT
$$
\begin{equation}\label{eq:psiintermed}
+
\frac{1}{2}\int_{-\infty}^{T_5}e^{-\varepsilon}\left[
\cos{\varepsilon}+\sin{\varepsilon}
-2x_0\left(\frac{1}{a}+\frac{1}{x}\right)\cos{\varepsilon}
+\frac{2x_0^2}{ax}\left(\cos{\varepsilon}-\sin{\varepsilon}\right)
\right]\,dT\,.
\end{equation}

\subsection{Static solution for $a>x_0$}

For comparison with evolution codes, it is important to know the initial 
data for evolution. This follows from setting $\omega=0$ in Eq.~(\ref{eq:FDGF})
and using the potential in Eq.~(\ref{eq:TDPpot}). The static solution must
have the form
\begin{equation}\label{static1}
  \Psi=\left\{
  \begin{array}{ll}
    \alpha/x &\mbox{for $x>a$}\\
    \beta/x+\xi x^2 &\mbox{for $x_0<x<a$}\\
    \delta &\mbox{for $x<x_0$}\\
  \end{array}
\right.\,.
\end{equation}
When the coefficients are matched at $x=a$ and $x=x_0$, so that the solutions
are continuous everywhere and have a unit jump in their derivative at $x=a$, 
the coefficients are found to be
\begin{equation}\label{static2}
  \alpha=-a^2/3-2x_0^3/3a\quad\quad
  \beta=-2x_0^3/3a\quad\quad
  \gamma=-1/3a\quad\quad
  \delta= -x_0^2/a\,.
\end{equation}

\subsection{Solutions for $a<x_0<x$}

\begin{figure}[h]
  \begin{center}
  \includegraphics[width=.2\textwidth ]{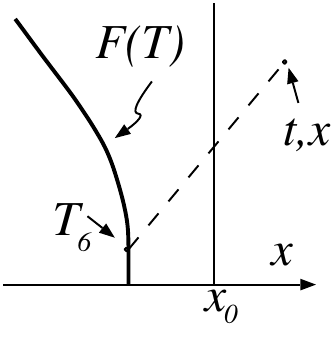}
  \caption{ A spacetime diagram showing the meaning of $T_6$. The bold
    line represents the trajectory $x=F(T)$ of the particle; the
    dashed line represents a radial characteristic.  }
  \label{fig:T6}
  \end{center}
  \end{figure}

Here the FDGF is
\begin{equation}
 {\cal G} =-\,
 \frac{i}{2}\frac{e^{-i\omega(a-x)}\omega}{(\omega-\omega_1)(\omega-\omega_2)}
 \left(1+\frac{i}{\omega x}\right)\,.
\end{equation}
We define $T_6$ and $\theta$ by 
\begin{equation}\label{T6thetadef}
  T_6-F(T_6)=t-x\quad\quad\quad \theta=(t-T+a-x)/2x_0\,,
\end{equation}
and illustrate them in Fig.~\ref{fig:T6}.

If the particle has {always} been moving in the $a<x_0$ region, then
the Green function gives the solution
\begin{equation}\label{eq:altx_outer}
  \Psi=\frac{x_0}{2}\,\int_{-\infty}^{T_6}e^{-\theta}
\left[\frac{1}{x_0}\left(\sin\theta-\cos\theta\right)-\frac{2}{x}\sin\theta\right]\,dT\,.
\end{equation}
In the case that mainly interests us, the particle will start at a
position greater than $x_0$ and cross $x_0$ at some time $T_{\rm
  cross}$.  We will need to calculate $\Psi$ for values of retarded
time $u$ both less than and greater than $u_{\rm crit}=T_{\rm
  cross}-x_0$. For $u<u_{\rm crit}$ we find $\Psi$ from
Eq.~(\ref{psixgtra2}). For $u>u_{\rm crit}$ we need
Eq.~(\ref{eq:altx_outer}), but the lower limit of integration must be
changed from $-\infty$, to $T_{\rm cross}$, since the $a<x_0$ Green
function only applies for $T>T_{\rm cross}$. To this integral must be
added the contributions for $T<T_{\rm cross}$. This comes from
Eq.~(\ref{psixgtra2}) adapted to $u>u_{\rm crit}$.

To understand how to adapt Eq.~(\ref{psixgtra2}), note that when
$u=u_{\rm crit}$ both $T_1$ and $T_2$ are equal to $T_{\rm cross}$.
Thus for $u>u_{\rm crit}$, the case we are interested in here, both
$T_1$ and $T_2$ are greater than $T_{\rm cross}$. But the integration
over this range of $T$ must not exceed $T=T_{\rm cross}$, since here
the Green function is only applicable when $F(T)>x_0$. Thus, both
$T_1$ and $T_2$ must be replaced by $T_{\rm cross}$ in
Eq.~(\ref{psixgtra2}). With this change, the first integral is missing
from Eq.~(\ref{psixgtra2}) and the second integration extends only to
$T_{\rm cross}$.  The complete expression then is
\begin{displaymath}
  \Psi=\frac{x_0}{2}\,\int_{T_{\rm cross}}^{T_6}e^{-\theta}
  \left[\frac{1}{x_0}\left(\sin\theta-\cos\theta\right)-\frac{2}{x}\sin\theta\right]\,dT
\end{displaymath}
\begin{equation}
    -\,\frac{1}{2}\,\int_{-\infty}^{T_{\rm cross}}
\,e^{-\gamma}\left[
-(\cos{\gamma}+\sin{\gamma})+2x_0\left(\frac{1}{a}+\frac{1}{x}\right)\cos{\gamma}
-\frac{2x_0^2}{ax}\left(\cos{\gamma}-\sin{\gamma}\right)
\right]\,dT\,.
\end{equation}

In the first integral, representing the contribution after the
particle has passed inward through $x_0$, the definition of $T_6$ and
$\theta$ are, as before, those given in Eq.~(\ref{T6thetadef}).  In
the second integral, representing the contribution before the particle
has crossed, the definitions of $\gamma$ is the same as that given for
$\xi$ in Eq.~(\ref{gammadef}).

\subsection{Solution for innner region $x<a$}
\begin{figure}[h]
  \begin{center}
  \includegraphics[width=.25\textwidth ]{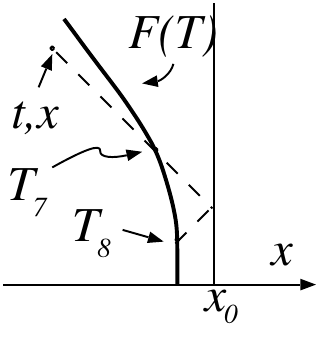}
  \caption{ A spacetime diagram showing the meaning of $T_7$ and
    $T_8$. The bold line represents the trajectory of the particle
    $x=F(T)$; the dashed lines are radial characteristics.  }
  \label{fig:T7T8}
  \end{center}
  \end{figure}

Here the Green function is
\begin{equation}
{\cal G}=-\,\frac{i}{2\omega}\,e^{-i\omega (x-a)}
+ \frac{i}{2\omega}\,e^{-i\omega(x+a-2x_0)}
+\frac{e^{-i\omega(x+a-2x_0)}}{2x_0(\omega-\omega_1)(\omega-\omega_2)}\,.
\end{equation}
We define
$T_7$ and $T_8$ as follows, and illustrate them in Fig.~\ref{fig:T7T8},
\begin{equation}
T_7+F(T_7)=t+x\quad\quad
T_8-F(T_8)=t+x-2x_0\,.  
\end{equation}

With $\varphi$ defined as
\begin{equation}
  \varphi=(t-T+x+a-2x_0)/2x_0\,,
\end{equation}
the solution for $\Psi$ becomes
\begin{equation}\label{eq:altxinner}
\Psi=-\frac{1}{2}\int_{T_8}^{T_7}\,dt-\int_{-\infty}^{T_8}
e^{-\phi}\sin\phi dT\,.
\end{equation}

\subsection{Solution for intermediate region $x_0>x>a$}
\begin{figure}[h]
  \begin{center}
  \includegraphics[width=.25\textwidth ]{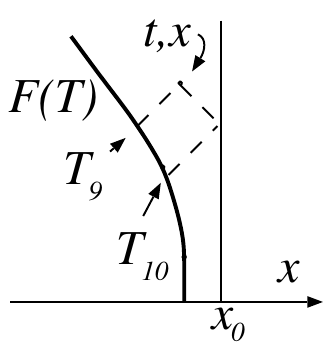}
  \caption{
A spacetime diagram showing the meaning of $T_9$ and $T_{10}$. The bold line
represents the trajectory of the particle $x=F(T)$; the dashed lines are 
radial
characteristics.
}
  \label{fig:T9T10}
  \end{center}
  \end{figure}

Here the Green function is
\begin{equation}
  {\cal G}=-\,i\frac{e^{-i\omega(a-x)}}{2\omega}
  -\,i\frac{e^{-i\omega(x+a-2x_0)}}{4\omega
    x_0^2(\omega-\omega_1)(\omega-\omega_2)}\,.
\end{equation}
We now define $\varphi$ and the times $T_9$, $T_{10}$ illustrated in
Fig.~\ref{fig:T9T10}.
\begin{equation}
  T_9-F(T_9)=t-x\quad\quad T_{10} -F(T_{10})=t+x-2x_0\quad\quad 
\varphi=(t-T+x+a-2x_0)/2x_0\,.
\end{equation}
With these we arrive at 
\begin{equation}\label{eq:intermed2}
  \Psi=-\,\frac{1}{2}\int_{T_{10}}^{T_9}dT
-\frac{1}{2}\int_{-\infty}^{T_{10}}e^{-\varphi}
\left(\cos\varphi+\sin\varphi\right)\,dT\,.
\end{equation}

\subsection{Static solution for $a<x_0$}

Let a unit particle be stationary at $x=a<x_0$.  We decompose the
static solution as follows:
\begin{equation}
  \Psi=\left\{
  \begin{array}{ll}
    \alpha/x &\mbox{for $x>x_0$}\\
    \beta+\gamma x &\mbox{for $a<x<x_0$}\\
    \delta &\mbox{for $x<a$}\\
  \end{array}
\right.\,.
\end{equation}
The conditions at $x=x_0$ are continuity and smoothness; the
conditions at $x=a$ are continuity and a unit jump in $d\Psi/dx$. The
solutions are easily found to be
\begin{equation}\label{aLTx0static}
  \Psi=\left\{
  \begin{array}{ll}
    -x_0^2/x &\mbox{for $x>x_0$}\\
    -2x_0+ x &\mbox{for $a<x<x_0$}\\
    -2x_0+a &\mbox{for $x<a$}\\
  \end{array}
\right.\,.
\end{equation}

\pagebreak

\end{document}